\documentclass[longauth]{aa}
\usepackage[varg]{txfonts}

\usepackage[T1]{fontenc}
\usepackage{ae,aecompl}
\usepackage{multirow}

\usepackage{graphicx}	
\usepackage{amsmath}	
\usepackage{amssymb}	
\usepackage{color}
\usepackage{hyperref}

\usepackage{array}
\newcolumntype{L}[1]{>{\raggedright\let\newline\\\arraybackslash\hspace{0pt}}m{#1}}
\newcolumntype{C}[1]{>{\centering\let\newline\\\arraybackslash\hspace{0pt}}m{#1}}
\newcolumntype{R}[1]{>{\raggedleft\let\newline\\\arraybackslash\hspace{0pt}}m{#1}}

\begin{document}


\title{The LOFAR Two-metre Sky Survey Deep Fields - Data Release 1} 

\subtitle{IV. Photometric redshifts and stellar masses\thanks{The redshift catalogues are available online at \href{https://lofar-surveys.org/}{https://lofar-surveys.org/}, as part of this data release.}$^{,}$\thanks{Catalogues are available in electronic form at the CDS via anonymous ftp to \href{cdsarc.u-strasbg.fr (130.79.128.5)}{cdsarc.u-strasbg.fr (130.79.128.5)} or via \href{http://cdsweb.u-strasbg.fr/cgi-bin/qcat?J/A+A/}{http://cdsweb.u-strasbg.fr/cgi-bin/qcat?J/A+A/}}} 
\authorrunning{Duncan et~al.}
\titlerunning{LoTSS Deep Fields: Photometric Redshifts}

\author{K. J. Duncan\inst{1,2}\and
R. Kondapally\inst{1} \and
M. J. I. Brown\inst{3} \and
M. Bonato\inst{4,5,6} \and
P.N. Best\inst{1}\and
H. J. A. R\"{o}ttgering\inst{2}\and
M. Bondi\inst{4}\and
R. A. A. Bowler\inst{7} \and
R. K. Cochrane\inst{8} \and
G. G\"{u}rkan\inst{9} \and
M. J. Hardcastle\inst{10} \and
M. J. Jarvis\inst{7,11} \and
M. Kunert-Bajraszewska\inst{12}\and
S. K. Leslie\inst{2}\and
K. Ma\l{}ek\inst{13,14} \and
L. K. Morabito\inst{15} \and
S. P. O'Sullivan\inst{16} \and
I. Prandoni\inst{6} \and
J. Sabater\inst{1} \and
T. W. Shimwell\inst{2,17} \and
D. J. B. Smith\inst{10} \and
L. Wang\inst{18,19} \and
A. Wo\l{}owska\inst{12}}

\institute{
SUPA, Institute for Astronomy, Royal Observatory, Blackford Hill, Edinburgh, EH9 3HJ, UK \and
Leiden Observatory, Leiden University, PO Box 9513, NL-2300 RA Leiden, The Netherlands \and
School of Physics and Astronomy, Monash University, Clayton, Victoria 3800, Australia\and 
INAF-Istituto di Radioastronomia, Via Gobetti 101, I-40129, Bologna, Italy\and
Italian ALMA Regional Centre, Via Gobetti 101, I-40129, Bologna, Italy\and
INAF-Osservatorio Astronomico di Padova, Vicolo dell'Osservatorio 5, I-35122, Padova, Italy\and
Astrophysics, Department of Physics, Keble Road, Oxford, OX1 3RH, UK\and
Harvard-Smithsonian Center for Astrophysics, 60 Garden St, Cambridge, MA 02138, USA\and
CSIRO Astronomy and Space Science, PO Box 1130, Bentley WA 6102, Australia\and
Centre for Astrophysics Research, University of Hertfordshire, College Lane, Hatfield AL10 9AB, UK\and
Department of Physics \& Astronomy, University of the Western Cape, Private Bag X17, Bellville, Cape Town, 7535, South Africa\and
Institute of Astronomy, Faculty of Physics, Astronomy and Informatics, NCU, Grudziadzka 5, 87-100 Toru\'n, Poland\and
National Centre for Nuclear Research, ul. Pasteura 7, 02-093 Warszawa, Poland\and 
Aix Marseille Univ. CNRS, CNES, LAM, Marseille, France\and
Centre for Extragalactic Astronomy, Department of Physics, Durham University, Durham, DH1 3LE, UK\and
School of Physical Sciences and Centre for Astrophysics \& Relativity, Dublin City University, Glasnevin, D09 W6Y4, Ireland\and
ASTRON, Netherlands Institute for Radio Astronomy, Oude Hoogeveensedijk 4, 7991 PD, Dwingeloo, The Netherlands\and 
SRON Netherlands Institute for Space Research, Landleven 12, 9747 AD, Groningen, The Netherlands\and
Kapteyn Astronomical Institute, University of Groningen, Postbus 800, 9700 AV Groningen, the Netherlands
}

\date{}

\abstract{The Low Frequency Array (LOFAR) Two-metre Sky Survey (LoTSS) is a sensitive, high-resolution 120-168\,MHz survey split across multiple tiers over the northern sky.
The first LoTSS Deep Fields data release consists of deep radio continuum imaging at 150 MHz of the Bo\"{o}tes, European Large Area Infrared Space Observatory Survey-North 1 (ELAIS-N1), and Lockman Hole fields, down to rms sensitivities of $\sim$32, 20, and 22 $\mu$Jy beam$^{-1}$, respectively.
In this paper we present consistent photometric redshift (photo-$z$) estimates for the optical source catalogues in all three fields - totalling over 7 million sources ($\sim5$ million after limiting to regions with the best photometric coverage).
Our photo-$z$ estimation uses a hybrid methodology that combines template fitting and machine learning and is optimised to produce the best possible performance for the radio continuum selected sources and the wider optical source population. 
Comparing our results with spectroscopic redshift samples, we find a robust scatter ranging from 1.6 to 2\% for galaxies and 6.4 to 7\% for identified optical, infrared, or X-ray selected active galactic nuclei (AGN).
Our estimated outlier fractions ($\left |  z_{\textup{phot}} - z_{\textup{spec}} \right | / (1+z_{\text{spec}}) > 0.15$) for the corresponding subsets range from 1.5 to 1.8\% and 18 to 22\%, respectively.
Replicating trends seen in analyses of previous wide-area radio surveys, we find no strong trend in photo-$z$ quality as a function of radio luminosity for a fixed redshift.
We exploit the broad wavelength coverage available within each field to produce galaxy stellar mass estimates for all optical sources at $z < 1.5$.
Stellar mass functions derived for each field are used to validate our mass estimates, with the resulting estimates in good agreement between each field and with published results from the literature.} 

\keywords{galaxies: distances and redshifts -- galaxies: active  -- radio continuum: galaxies}
\maketitle 

\defcitealias{Duncan:2017wu}{D18a}
\defcitealias{Duncan:2017ul}{D18b}
\defcitealias{Shimwell:2018to}{DR1-I}
\defcitealias{Williams:2018us}{DR1-II}
\defcitealias{2019A&A...622A...3D}{D19}
\defcitealias{Duncan:2017wu,Duncan:2017ul}{D18a,b}
\defcitealias{Kondapally2020}{Paper III}

\section{Introduction}
Combining extremely high sensitivity with a wide field-of-view, the Low Frequency Array \citep[LOFAR;][]{vanHaarlem:2013gi} offers an unprecedented capability for performing large statistical surveys of the radio sky.
The LOFAR Surveys Key Science Project \citep{2010iska.meetE..50R} is undertaking a set of tiered surveys over the northern sky at 120-168\,MHz.
The first release of data from the all-sky LOFAR Two-metre Sky Survey \citep[LoTSS;][]{Shimwell:2017ch} presented 424deg$^{2}$ to an average sensitivity of 71$\mu$Jy beam$^{-1}$ at 150\,MHz\footnote{Formally, the central frequency of the LoTSS Deep Fields data is 144\,MHz in Bo\"{o}tes and Lockman Hole, and 146\,MHz in ELAIS-N1. However, throughout this paper we will refer to the LoTSS frequency colloquially as 150 MHz.}.
Complementary to the wide area LoTSS data, the LoTSS Deep Fields First Data Release (LoTSS DR1) reaches radio continuum sensitivities comparable to or deeper than the deepest surveys currently available, and over orders of magnitude wider areas ($>50$ deg$^{2}$; see \citet{Tasse:2020tr} and \citet{Sabater:2020ur} - Paper I and Paper II, respectively).
Located in some of the best-studied northern extragalactic survey fields - Bo\"{o}tes, European Large Area Infrared Space Observatory Survey-North 1 (ELAIS-N1, or EN1 hereafter), and the Lockman Hole (LH) - the LoTSS Deep Field data reach a current rms sensitivity of $\sim$32, 20, and 22 $\mu$Jy beam$^{-1}$ at 150 MHz, respectively, sufficient to detect radio-quiet AGN and extremely star-forming galaxies out to the highest redshifts ($z > 5$).

Extracting the maximum scientific value from the LOFAR radio continuum observations requires robust identification of the host galaxies of radio sources, alongside knowledge of the source redshifts, to extract intrinsic physical properties for both the radio sources (e.g. physical size, luminosity) and their host galaxies.
By design, the LOFAR deep fields are located in regions of the sky that contain extensive ancillary imaging data, from ultraviolet (UV) all the way to far-infrared (FIR).
Kondapally et al. (2020; hereafter \citetalias{Kondapally2020}) present new multi-wavelength optical to mid-IR photometry catalogues for all three fields alongside careful cross-identification with the LOFAR radio source population (with host identifications for $\gtrsim97\%$ of radio sources).
However, the wide range of intrinsic host properties in radio continuum selected samples, including both active galactic nuclei (AGN) and extreme star-forming galaxies, means that even with extensive ancillary data, deriving reliable photometric redshifts (photo-$z$s) can be non-trivial \citep[][]{2019PASP..131j8004N}.

In \citet[][hereafter D18a]{Duncan:2017wu}, we demonstrated that photo-$z$ estimates for radio continuum sources obtained by combining multiple template-fitting estimates can be both more precise and more reliable than those using just one single template library.
However, there remained key subsets of the AGN population for which template-based photo-$z$s could not produce satisfactory results (e.g. IR or X-ray selected AGN at $1 < z < 3$).

In \citet[][hereafter D18b]{Duncan:2017ul}, we built upon the previous template-fitting method of \citetalias{Duncan:2017wu} by incorporating additional machine-learning-based photo-$z$ estimates trained for different AGN subsets.
When combined together within a hierarchical Bayesian (HB) combination framework, the resulting consensus photo-$z$ estimates improve on the performance of either individual method \citep[specifically template-fitting or machine learning, see also][for other successful approaches]{2017MNRAS.466.2039C,2018A&A...619A..14F}.
This combined `hybrid' approach was successfully applied to shallow optical and mid-IR photometry over very wide fields as part of LoTSS DR1 \citep[][]{Shimwell:2018to,2019A&A...622A...2W}.
The resulting photo-$z$s - presented in \citet[][hereafter D19]{2019A&A...622A...3D} - provided precise and reliable photo-$z$s ($\approx 3\%$ scatter and $<2\%$ outlier fraction; OLF) out to $z\sim0.8$ for LOFAR radio sources with host-dominated optical-infrared spectral energy distributions (SEDs; specifically, sources for which there is no strong evidence for a dominant AGN contribution to the SED at X-ray, optical, or IR wavelengths).
Beyond $z\sim0.8$, the shallow optical data leads to incomplete samples and highly uncertain redshift estimates for this population in LoTSS DR1.
Thanks largely to the extensive training samples available for machine learning estimates, reliable photo-$z$ estimates for quasar-like sources in \citetalias{2019A&A...622A...3D} extended out to $z\lesssim 3$.

In the case of the Bo\"{o}tes field, a test field for both \citetalias{Duncan:2017wu} and \citetalias{Duncan:2017ul}, we have already demonstrated that our hybrid approach can produce high quality photo-$z$s across a broad range of source types and redshifts.
However, Bo\"{o}tes is currently unique among the LoTSS Deep Fields with respect to its larger sample of high quality spectroscopic redshifts both for calibration of template fits and for training of machine learning photo-$z$ estimates.
Neither the EN1 nor LH fields contain the representative samples of AGN spectroscopic redshifts necessary to apply the hybrid method of \citetalias{Duncan:2017ul} with the same effectiveness as achieved in Bo\"{o}tes.
Until large samples of spectroscopic redshifts provided by the forthcoming WEAVE-LOFAR survey \citep{Smith:2016vw}, a modified approach is therefore required if we wish to maximise the quality of photo-$z$s across all three LOFAR Deep Fields and fully exploit the extraordinary sample of faint radio continuum sources they provide. 

In this paper we present photometric analysis of the matched aperture multi-wavelength optical to mid-IR data in each of the three LoTSS Deep Fields.
The primary aim of the paper is to provide optimal photo-$z$ estimates for all sources in the fields to enable the exploitation of the deep radio continuum observations.
Additionally, given that all three fields have deep near-infrared (NIR) or mid-infrared (mid-IR) photometry required for robust SED fitting, we provide stellar mass estimates for the subset of the optically detected sources for which we can make reliable measurements - enabling a wide range of science and providing a valuable reference sample for studies of the faint LOFAR radio population.  

%
The remaining sections of this paper are set out as follows.
In Section~\ref{sec:data} we summarise the properties of the multi-wavelength photometry and spectroscopic samples used for photo-$z$ analysis in this study.
Next, in Section~\ref{sec:photoz} we outline the key changes to the hybrid photo-$z$ method used in this analysis.
In Section~\ref{sec:results} we analyse the precision and accuracy of the resulting photo-$z$s across all three fields as a function of redshift, magnitude, source type and radio luminosity.
Section~\ref{sec:stellarmasses} then presents the method used for deriving stellar mass estimates for the optical sample, along with derivation of mass completeness limits and tests demonstrating their overall quality and limitations.
In Section~\ref{sec:radioproperties} we then briefly explore the physical properties of the LoTSS Deep Fields radio source population based on our derived redshifts and stellar masses.
Finally, Section~\ref{sec:summary} presents a summary of our work.
Throughout this paper, all magnitudes are quoted in the AB system \citep{1983ApJ...266..713O} unless otherwise stated. We also assume a $\Lambda$ Cold Dark Matter cosmology with $H_{0} = 70$ km\,s$^{-1}$\,Mpc$^{-1}$, $\Omega_{m}=0.3$ and $\Omega_{\Lambda}=0.7$.

\section{Data}\label{sec:data}
\subsection{Photometry}
Optical to mid-IR imaging, photometry and radio source identification for each of the LOFAR deep fields are presented in \citetalias{Kondapally2020}, to which we refer the reader for full details.
In summary, available deep photometry in the EN1 and Lockman Hole fields has been mosaicked onto a common pixel and flux scale.
For each field, forced aperture photometry in all bands has then been performed based on detections in two stacked $\chi^2$ S/N images (optical to NIR and mid-IR) and the resulting sets of catalogues have been merged to produce a consistent multi-wavelength catalog.
In the Bo\"{o}tes field, existing forced aperture photometry catalogues based on detections in $I$ and \emph{Spitzer} Infrared Array Camera \citep[IRAC;][]{Fazio:2004eb} 4.5$\mu$m images have been merged following the same procedure as used for the other two fields.
Aperture corrections based on either curve of growth analysis for sources in the field (EN1 and LH) or the analytic point spread function (PSF; Bo\"{o}tes) have then been calculated to provide total flux estimates in each filter.

\begin{figure}[h!]
\centering
 \includegraphics[width=0.95\columnwidth]{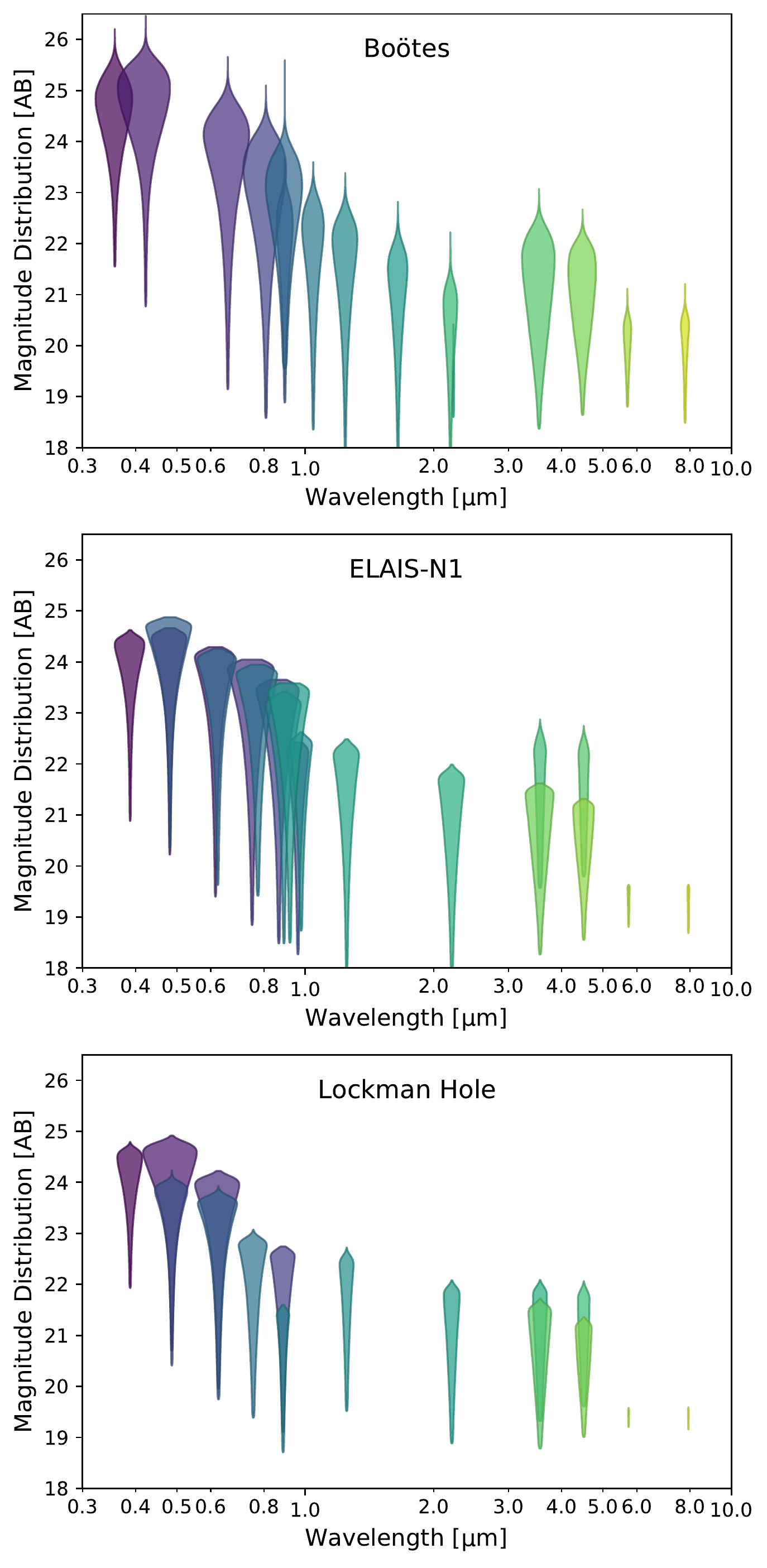}
 \caption{Illustration of the wavelength coverage and depths for the photometric datasets used for photo-$z$ estimation in the three LoTSS Deep Fields. The violin plot for each filter shows the distribution of magnitudes for $5\sigma$ sources. The widths of the violin plots are indicative of the area available at a given depth within each field and are scaled by the fraction of sources with $5\sigma$ detections compared to the total size of the catalog. For example, the deeper \emph{Spitzer} IRAC photometry of the smaller SERVS regions (at 3.6 and 4.5$\mu$m) are evident in the EN1 and LH plots.}
 \label{fig:photom_depths_violin}
\end{figure}

For all three fields we make use of 3$\arcsec$ apertures for optical to NIR bands and 4$\arcsec$ apertures for IRAC (due to the lower resolution in the IRAC imaging).
As described in \citetalias{Kondapally2020}, the galactic extinction ($E(B-V)$) for each source has been calculated based on the Milky Way extinction map of \citet{1998ApJ...500..525S}, queried using the \textsf{dustmaps} Python package \citep{dustmaps2018}.
Filter-dependent extinction factors are then calculated by convolving the respective filter response curves with the Milky Way dust extinction law of \citet{1999PASP..111...63F}.
In Figure~\ref{fig:photom_depths_violin} we visually summarise the wavelength coverage and sensitivity of the resulting photometry catalogues used for the photo-$z$ analysis in this paper. 

In addition to the processing steps above that are presented in \citetalias{Kondapally2020}, for the photo-$z$ estimation and SED fitting work in this paper we include further processing as follows.
Due to the combination of large areas, the number of unique bands and depth of the photometry, small numbers of spurious datapoints are inevitable.
Therefore, as a final additional step before photo-$z$ estimation, we automatically filter the photometric datasets for spurious datapoints.
Specifically, we filter for unphysical colours or extreme datapoints that are likely to be caused by artefacts such as stellar diffraction spikes, cosmic rays, cross-talk etc.
Measurements that lie $>2.5$ mag above or $<1$ mag below those at \emph{both} adjacent filters at shorter and longer wavelengths, indicative of extreme excess or deficit, are excluded.
The large magnitude cuts are chosen to not exclude strong emission or absorption line sources. 
Similarly, measurements that result in a $>5$ magnitude jump or drop in colour between consecutive filters are also excluded.
These cuts are designed to be conservative, excluding clearly unphysical colours whilst not affecting large but physical colours such as Lyman break or strong emission line features. 
We note that the full optical catalogues released in \citetalias{Kondapally2020} have not been processed in this way as users may wish to apply different cuts.
However, the exact photometric catalogues used for photo-$z$ estimation (including outlier identification) are available through the LOFAR Surveys Data Release site\footnote{\hyperlink{https://lofar-surveys.org/releases.html}{https://lofar-surveys.org/releases.html}}.

There likely remain a number of spurious measurements for which more detailed analysis would be required for identification \citep[e.g. iterative fits excluding individual photometry points, c.f. ][]{Chung:2014it}.
However, given the large number of available bands and the overall high quality of imaging available in these fields, we do not expect that our photo-$z$ estimates (or science results derived from them) are significantly affected by any remaining anomalous datapoints. 

\subsection{Multi-wavelength classifications}\label{sec:mw_class}

We broadly classify all sources in the photometric samples using the following additional criteria:

\emph{Infrared AGN} are identified using the \emph{Spitzer} IR colour and and monotonically increasing mid-IR SED criteria presented by \citet{Donley:2012ji}. As we are primarily concerned with identifying only the robustly selected AGN sources, we also require $>5\sigma$ detections in all four IRAC bands.
	For the subset of the optical catalogues which satisfy our primary quality cuts, we find 0.7, 0.3 and 0.3\% of the catalogues are found to satisfy the IR AGN criteria in Bo\"{o}tes, EN1, and LH, respectively.
	
\emph{Optical AGN:} were identified through cross-matching the optical catalogue with the Million Quasar catalogue compilation of optical AGN, which is primarily based on Sloan Digital Sky Survey \citep[SDSS;][]{2015ApJS..219...12A} and other literature catalogues \citep{2015PASA...32...10F}. 
Additionally, sources flagged as AGN based on spectroscopic observations are included in the optical AGN sample.
	For the subset of the optical catalogues which satisfy our primary quality cuts, we find consistently that $\approx 0.1\%$ of the optical sources in all three fields satisfy the optical AGN selection criteria.
	
\emph{X-ray AGN:} In the Bo\"{o}tes field where deep X-ray observations are present over a large area, X-ray AGN were identified by cross-matching the positions of sources in our catalogue with the X-B\"{o}otes \emph{Chandra} survey of NDWFS \citep{Kenter:2005gj}.
We calculate the X-ray-to-optical flux ratio, $X/O = \log_{10}(f_{X}/f_{\textup{opt}})$, based on the $I$ band magnitude following \citet{Brand:2006iv}
For a source to be selected as an X-ray AGN, we require that an X-ray source has $X/O > -1$ or an X-ray hardness ratio $> 0.8$ \citep{2004AJ....128.2048B}.
In total, we find 2811 X-ray AGN in Bo\"{o}tes based on these criteria.
In EN1 and LH, bright X-ray sources were identified based on the Second \emph{ROSAT} All-Sky Survey \citep[2RXS;][]{2016A&A...588A.103B} and the \emph{XMM-Newton} Slew Survey (XMMSL2)\footnote{\hyperlink{https://www.cosmos.esa.int/web/xmm-newton/xmmsl2-ug}{https://www.cosmos.esa.int/web/xmm-newton/xmmsl2-ug}}, as in LoTSS DR1.
X-ray sources were matched to their IR counterparts using the published AllWISE \citep[Wide-field Infrared Survey Explorer, WISE;][]{Wright:2010in} cross-matches of \citet{Salvato:2017gj}, with sources then matched to the deep fields photometric dataset using the corresponding AllWISE source positions. 

We note here that as in previous works, these broad sample selections are designed to identify clear AGN dominated SEDs for the purposes of optimising the photo-$z$ analysis.
The classifications are not intended to be complete samples of the AGN population within the fields.
Subsequent studies will combine the multi-wavelength photometry with radio and FIR information to provide robust source classifications for the LOFAR detected population (see Best et al. 2020; Paper V).

\subsection{Spectroscopic redshift samples}
 Spectroscopic redshifts for sources in Bo\"{o}tes are taken from a compilation of observations within the field comprising primarily the results of the AGN and Galaxy Evolution Survey \citep[AGES;][]{Kochanek:jy} spectroscopic sample, with additional redshifts provided by a large number of smaller surveys in the field including \citet{2012ApJ...758L..31L,2013ApJ...771...25L,2014ApJ...796..126L}, \citet{2012ApJ...753..164S}, \citet{2012ApJ...756..115Z,2013ApJ...779..137Z} and \citet{2016ApJ...823...11D}.
 Included in this sample are a number of proprietary redshifts used for spectroscopic redshift training only (M. Brown, \emph{private communication}).
 In the release catalogues provided in this paper, we incorporate publicly available spectroscopic redshifts compiled as part of the Herschel Extragalactic Legacy Project (HELP, PI: S. Oliver)\footnote{\hyperlink{http://hedam.lam.fr/HELP/}{http://hedam.lam.fr/HELP/}}.

Our spectroscopic redshift samples for the EN1 and LH fields are also based on the HELP compilations.
The majority of spectroscopic redshifts in these compilations originate from the SDSS spectroscopic sample \citep{2015ApJS..219...12A}, with additional data from a number of smaller spectroscopic follow-up campaigns \citep[namely][]{2007A&A...467..565B, 2007MNRAS.376..479S, 2007MNRAS.379.1343S, 2013ApJS..208...24L}.

\begin{table}
\centering
\caption{Total spectroscopic redshift samples available for photo-$z$ training and/or validation and the size of the corresponding photometric catalogue used in this analysis. We note that these numbers represent the total contained within the full catalogues and do not account for any cuts applied to the catalogues for analysis (e.g. flagging and restrictions to regions with specific photometric coverage).}\label{tab:specz_samples}
\begin{tabular}{lccc}
    Field & $N_{\textup{spec}-z}$ (Galaxies) & $N_{\textup{spec}-z}$ (AGN) & $N_{\textup{Total}}$\\
    \hline
  Bo\"{o}tes  & 19\,143 & 2\,714 & 2\,214\,329\\
  EN1 & 3\,419 & 593 & 2\,105\,993 \\
  LH & 4\,787 & 1\,182 & 3\,041\,794 
  \end{tabular}
\end{table}

\begin{figure}[h!]
\centering
 \includegraphics[width=0.95\columnwidth]{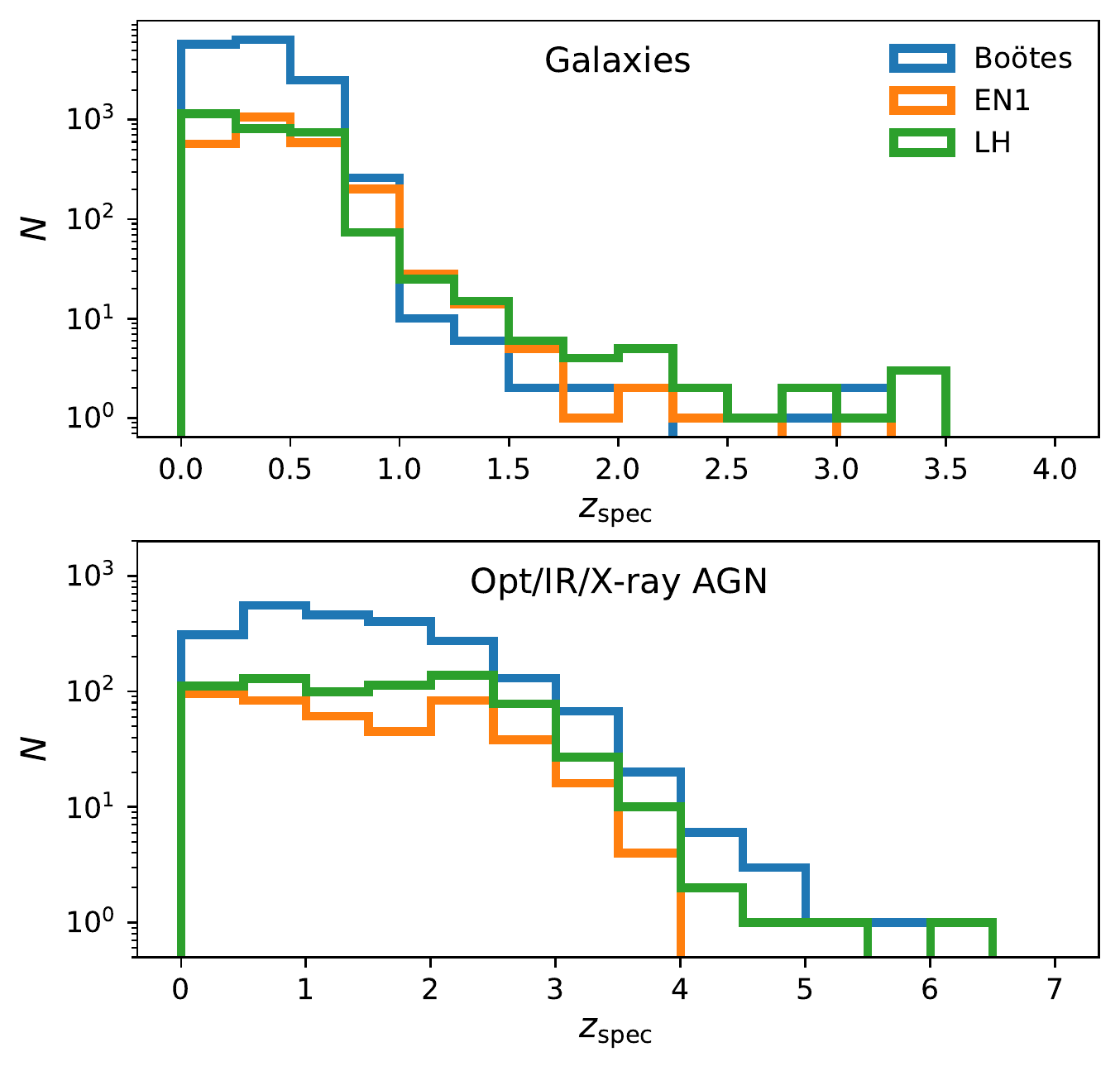}
 \caption{Redshift distributions for the spectroscopic redshift training and test samples available in each of the fields (see Table~\ref{tab:specz_samples} for total numbers). While Bo\"{o}tes has a much greater number of sources with $z_{\text{spec}}$ available, these are largely limited to $z < 1$ for galaxies but extend over a greater redshift range for the known AGN.}
 \label{fig:specz_zdist}
\end{figure}

\begin{figure}[h!]
\centering
 \includegraphics[width=0.95\columnwidth]{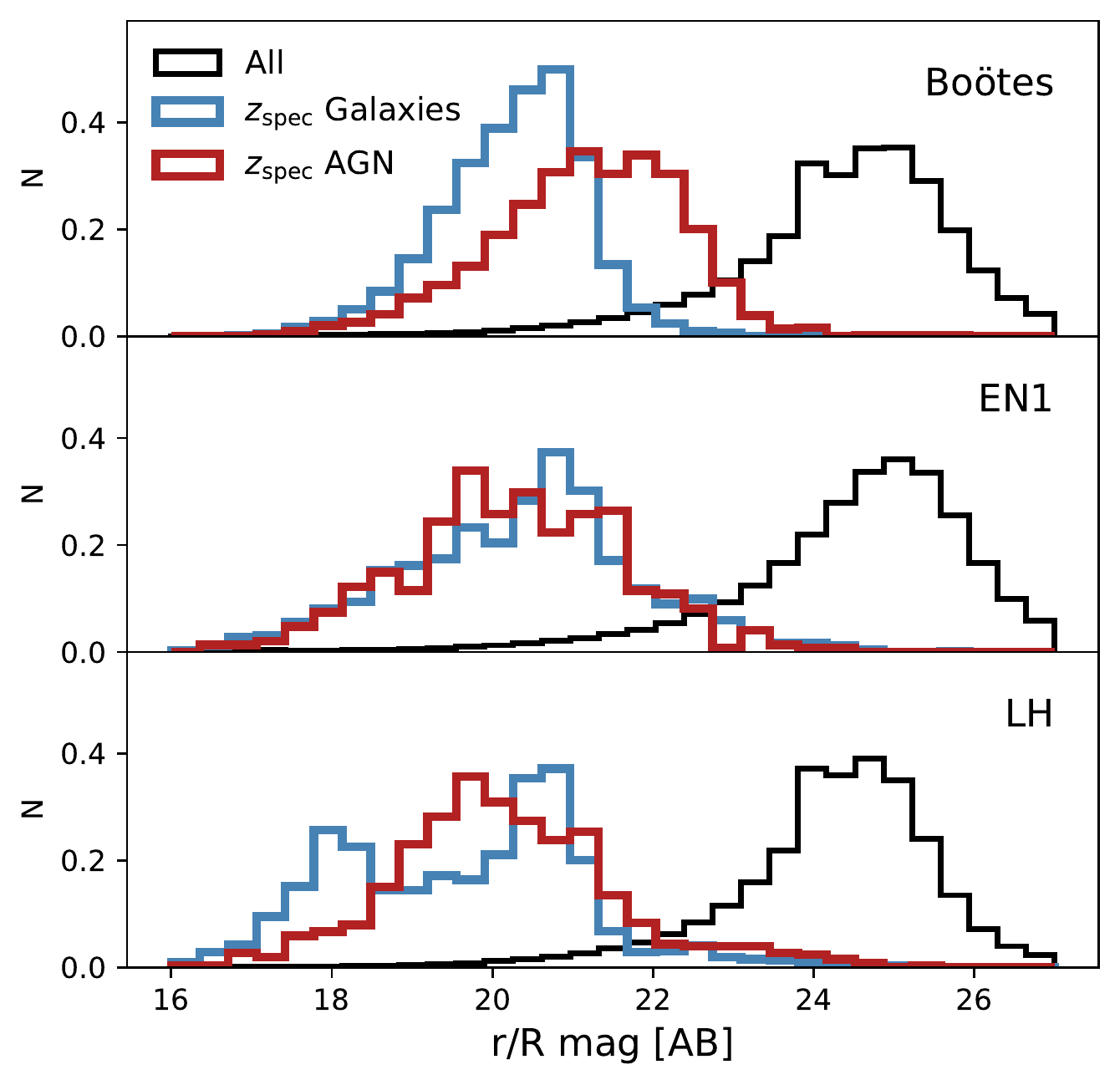}
 \caption{Normalised optical magnitude distributions for the galaxy (blue histogram) and AGN (red histrogram) spectroscopic redshift samples in comparison to the full parent photometric sample. For EN1 and LH we plot the $r$-band magnitude distribution, while for Bo\"{o}tes we show the nearest equivalent wavelength, $R$. The offset in the magnitude distributions highlights the importance of high quality template estimates in the regime where spectroscopic training samples are not available.}
 \label{fig:specz_magdist}
\end{figure}

For all three fields, the spectroscopic samples are matched to the optical photometry catalogues using a simple nearest neighbour match with a maximum radius of $1\arcsec$.
Table~\ref{tab:specz_samples} summarises the spectroscopic redshift samples available for each field and the corresponding redshift and magnitude distributions are presented in Fig.~\ref{fig:specz_zdist} and Fig.~\ref{fig:specz_magdist}, respectively.
The disparity in currently available spectroscopic samples between Bo\"{o}tes and the other two deep fields is evident, with much larger samples of low-redshift galaxies available and significant AGN samples that extend to higher redshift and fainter magnitudes than available in EN1 or LH.
However, we can see from Fig.~\ref{fig:specz_magdist} that the available spectroscopic sources are biased towards brighter magnitudes in all fields.
Nevertheless, a key goal of this work is to provide consistent photo-$z$ estimates across the three fields in all parameter space - ideally of comparable quality.

\section{Photometric redshift estimates}\label{sec:photoz}
We estimate photo-$z$s for the full matched aperture optical catalogues presented in \citetalias{Kondapally2020} following a modified version of the hybrid approach presented in \citetalias{Duncan:2017wu} and \citetalias{Duncan:2017ul} and applied to LoTSS DR1 in \citetalias{2019A&A...622A...3D}. 
For full details of the motivation behind the hybrid approach, as well as details regarding the HB combination method we refer the reader to those papers.
In this paper we focus solely on the modifications or changes to the method presented in \citetalias{2019A&A...622A...3D} that are specific to this work.

\subsection{Template fitting}\label{sec:method-templates}
Due to the limited availability of spectroscopic training samples in two of the deep fields (EN1 and Lockman Hole) and the scientific focus on higher redshift where spectroscopic redshifts will always be limited, the primary modifications to our hybrid photo-$z$ method are designed to maximise the accuracy and reliability of our template-fitting based estimates for the AGN population.
Our aim is to reduce the potential need for machine learning estimates and provide consensus photo-$z$s for AGN in EN1 and LH that are of comparable quality to those produced in Bo\"{o}tes \citepalias{Duncan:2017ul}.
For the template fitting photo-$z$ estimates, we use updated versions of all three template libraries that either improve the wavelength coverage or broaden the representation of different AGN SED types.
The three updated libraries now used for template fitting are:

\emph{Updated \textsc{Eazy} models:} In previous work, the `default' \textsc{Eazy} template set (version 1.3) was found to produce photo-$z$ estimates with the smallest scatter and OLF for star-forming or quiescent galaxy populations \citepalias{Duncan:2017wu}. 
    Here we make use of the new template set derived from the Flexible Stellar Population Synthesis code \citep[FSPS;][]{Conroy:2009ks,2010ascl.soft10043C}.
    While designed to reproduce the same representative combination of stellar emission dominated galaxy SEDs, the revised FSPS templates now include dust reprocessed emission in the mid-IR.
    The emission from dust extends the range of rest-frame wavelengths that can reliably be used to constrain the photo-$z$ estimate, which is particularly valuable in the deep field datasets that includes photometry out to 8\,$\mu$m.
    
\emph{Extended Atlas Library:} The second template set, the `Atlas of Empirical SEDs' \citet{Brown:2014jd} was recently extended to incorporate a wide range of the AGN and quasi-stellar object (QSO) populations not previously present in the original library \citep{2019MNRAS.489.3351B}. Additionally, the new Atlas of AGN SEDs incorporates a range of combinations of Seyfert type AGN spectra and different underlying host stellar populations (with varying AGN to host contributions). 
    As illustrated in \citet{2019MNRAS.489.3351B}, the inclusion of a fully representative range of AGN SEDs (and combinations of AGN + host galaxy) leads to significant improvement in the photo-$z$ statistics for AGN dominated galaxies - the area for which machine learning estimates were found to provide the greatest improvement in the hybrid method \citepalias{Duncan:2017ul}.
    
\emph{Revised `XMM-COSMOS' Team templates:} Finally, we also make use of a new iteration of the `XMM-COSMOS' galaxy and AGN SEDs \citep{Polletta:2007ha, Salvato:2008ef, Salvato:2011dq} as presented in \citet{2017ApJ...850...66A}. Optimised for significantly larger survey fields than previous iterations (comparable to the total LOFAR Deep Fields coverage), the implementation in \citet{2017ApJ...850...66A} includes more luminous quasar SED types than previous versions - as would be expected in the larger survey volumes than the deep pencil-beam surveys for which the \citet{Salvato:2011dq} library was optimised.

The implementation of all three libraries within our template fitting method is the same as their corresponding versions in \citetalias{Duncan:2017wu} and \citetalias{Duncan:2017ul}.
One key exception is the incorporation of new template library specific rest-frame model uncertainties within the fitting process that we outline below.

\subsubsection{Photometric zero-point offsets}
The inclusion of small magnitude offsets, or zero-point offsets, to the observed photometry of some datasets has been shown to improve photo-$z$ estimates from template fitting \citep[e.g. see][]{Dahlen:2013eu}. While typically small ($\lesssim10\%$), these additional offsets can often substantially reduce the overall scatter or OLF for photo-z estimates.
In \citet{Hildebrandt:2012du}, detailed comparison between the resulting photo-$z$ performance for different levels of photometry precision indicates that zero-point flux offsets serve largely to correct for PSF effects.
Given the nature of the photometry used in the work, we would therefore expect that the inclusion of zero-point offsets during template fitting will be beneficial.

Zero-point offsets for all template sets are derived for each photometric dataset following the method outlined in \citetalias{Duncan:2017wu}.
In summary, for 50\% of the spectroscopic redshift subset (with 50\% retained for validation/testing), the template set is fit to the observed photometry with the redshift fixed to the true redshift and the corresponding zeropoint offset is then calculated from the median offset between the observed and fitted flux values for sources with $S/N > 3$ in that band.

As previously seen for the Bo\"{o}tes photometry in \citetalias{Duncan:2017wu}, the inclusion of the zero-point offsets during template fitting leads to substantial improvement in photo-$z$ quality (as tested using the 50\% of spectroscopic sources not included in the derivation of zero-point offsets).
Similar improvements are observed for both EN1 and LH.
For reference, we present the derived zero-point offsets for all template library and field combinations in Appendix~\ref{app:zeropoints}.
Within a given field, we find good agreement between the largest zero-point corrections derived for the different template sets (e.g. the UKIRT Infrared Deep Sky Survey, UKIDSS, $K$-band data in EN1 and LH are consistently found to require a correction of $\sim10\%$).  
We conclude that for the datasets employed in this analysis, the zero-point corrections are likely correcting for small offsets in the photometry itself, rather than large systematic errors in the templates themselves (we note that the three different template sets encompass both model and empirically derived templates).

\subsubsection{Template model uncertainties}\label{sec:photoz-modelerrors}
Incorporating an estimate of the rest-frame model uncertainties during photo-$z$ template fitting has been shown to significantly improve the accuracy of the resulting estimates \citep{Brammer:2008gn, Dahlen:2013eu}.
As part of our revised photo-$z$ template fitting procedure we therefore estimate the rest-frame uncertainties for each template library from the spectroscopic training sample following the method outlined by \citet{Brammer:2008gn}.
For the Bo\"{o}tes field (chosen due to its large available spec-$z$ sample), for each template set we fit the templates to the spectroscopic sample (including the zero-point flux correction) while fixing the fit to the known redshift and then measuring the distribution of flux residuals.
\begin{figure}[h!]
\centering
 \includegraphics[width=0.95\columnwidth]{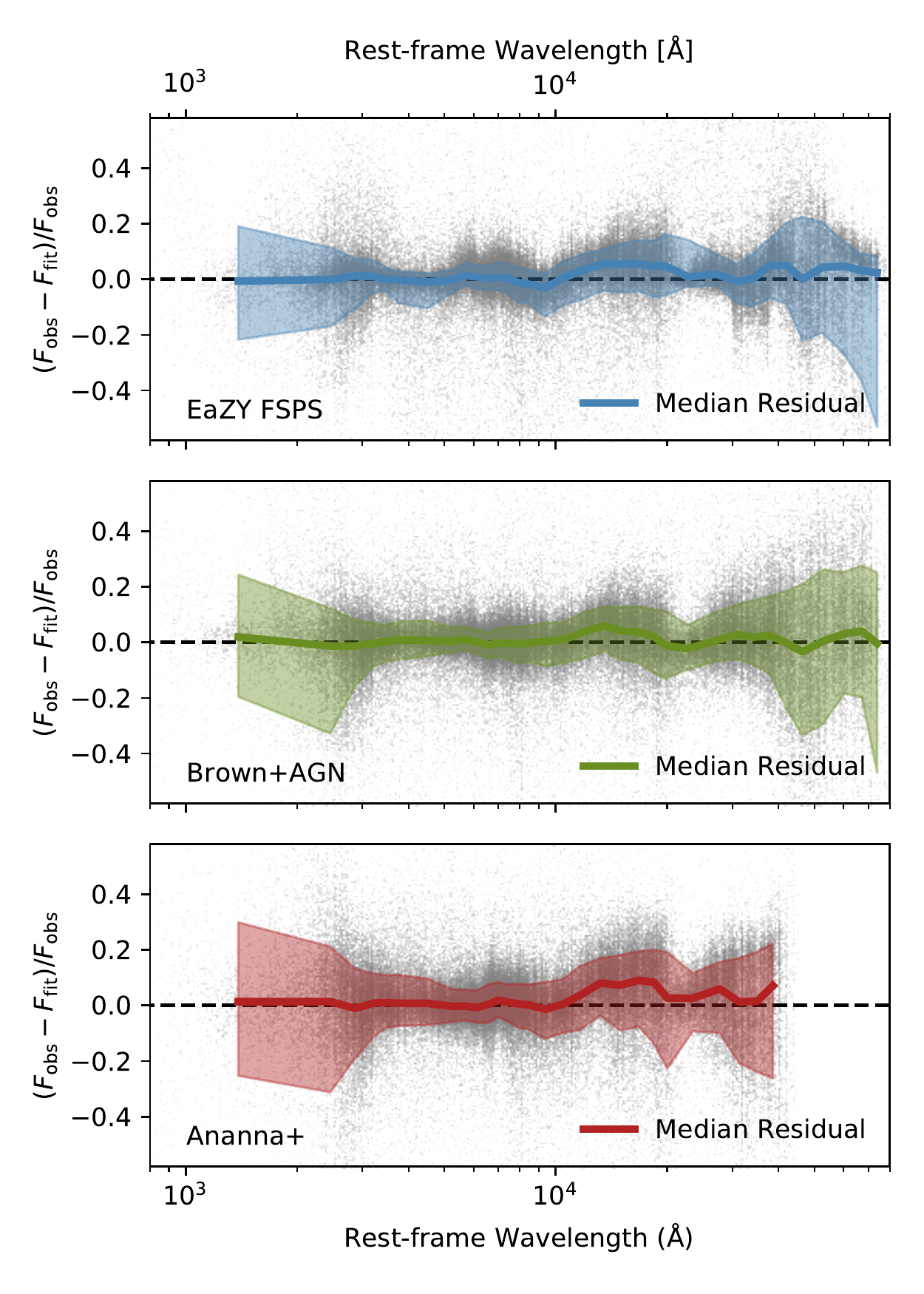}
 \caption{Rest-frame residuals for the template fits using each of the three SED libraries employed in this work. Background grey points correspond to individual data points (i.e. one for each fitted filter per source) while the solid coloured lines correspond to the median residual within a given wavelength bin and the shaded region corresponds to the 1$\sigma$ distribution (16 and 84th percentiles).}
 \label{fig:rf_residuals_bootes}
\end{figure}

\begin{figure}[h!]
\centering
 \includegraphics[width=0.9\columnwidth]{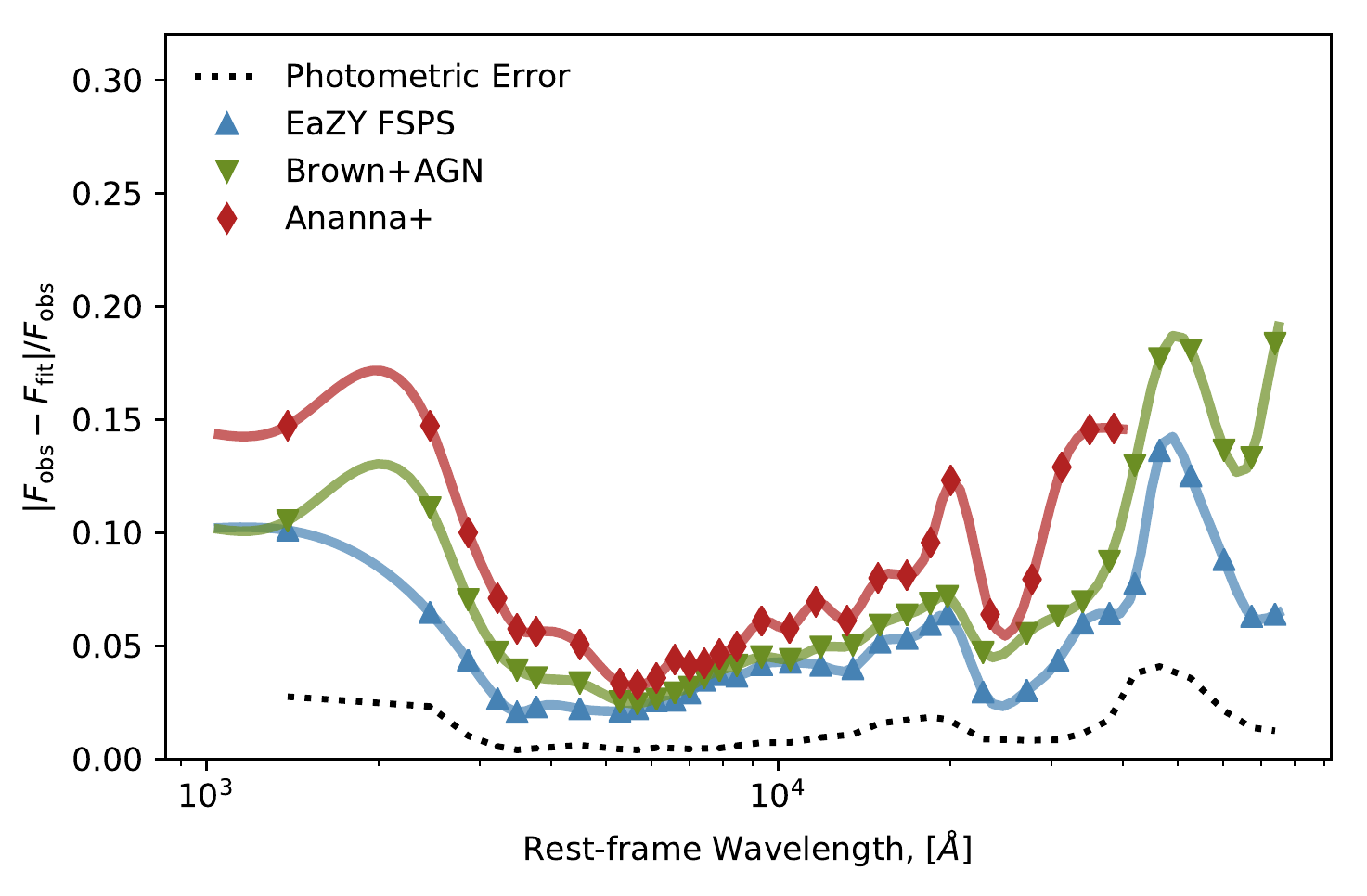}
 \caption{Rest-frame model uncertainties derived for each of the template libraries used during our analysis (solid coloured lines). The model uncertainties are calculated by subtracting in quadrature the scaled average fractional error (black dotted line) from the median absolute rest-frame residuals (coloured symbols). However, the contribution of the average error is negligible.}
 \label{fig:temp_err_bootes}
\end{figure}

The measured rest-frame residuals for each template library and the corresponding 1$\sigma$ ranges are illustrated in Fig.~\ref{fig:rf_residuals_bootes}.
The resulting template error functions for each library are then shown in Fig.~\ref{fig:temp_err_bootes}.
Consistent with previous measurements of the photo-$z$ template errors \citep{Brammer:2008gn}, we find that rest-frame UV and mid-IR wavelengths have the highest model uncertainties. 
We note however that model uncertainties estimated by this method are representative of the average uncertainty across the full template set given the spectroscopic sample available.
As demonstrated by \citet{2019MNRAS.489.3351B}, the model uncertainties for known AGN samples can be substantially larger at certain rest-frame wavelengths.
However, a full Bayesian template fitting framework that can account for variance associated with each individual template \citep[see e.g.][]{2019ApJ...881...80L} is not practical for this work.

In our implementation of the rest-frame model uncertainties for this study, we also identify and correct for a subtle but important statistical error in the version of the \citet{Brammer:2008gn} photo-$z$ code used.
Specifically, by default \textsc{Eazy} simply stores the best-fit $\chi^2$ at each step in the chosen redshift grid ($\hat{\chi}^{2}(z)$). 
The redshift likelihood is then typically taken as $\propto \exp{(-\hat{\chi}^{2}(z) / 2)}$ \citep[e.g.][]{Dahlen:2013eu, Finkelstein:2014ub}, which may then be convolved with a redshift prior (e.g. a magnitude prior).
However, when redshift-dependent model errors are included in the fitting, the simplifying assumption of dropping the normalisation in the Gaussian likelihood term is no longer valid since it is no longer a constant. 
This has the effect of the $P(z)$ becoming biased towards redshift ranges where the template error itself is maximised. For bright sources with strong colour features, this has minimal effect, but for fainter sources this can significantly bias the $P(z)$.
Correctly accounting for the redshift-dependent model errors is trivial, with the maximum likelihood at a given redshift, $\hat{\mathcal{L}}(z)$, being:
\begin{equation}
\hat{\mathcal{L}}(z) = \prod_{i}^{N} \frac{1}{\sqrt{2\pi\sigma_{i}(z)}} \exp \left ( - \frac{(f_{i,\text{obs}} - \hat{f}_{i,\text{t}}(z))^{2}}{2\sigma_{i}(z)^2} \right ),
\end{equation}
where for a given filter, $i$, $\sigma_{i}(z)$ is the total error (i.e. $\sigma_{i}(z)^{2} = \sigma_{i, \textup{obs}}^{2} + \sigma_{i, \textup{model}}^{2}(z)$), $f_{i,\text{obs}}$ the observed flux and $\hat{f}_{i,\text{t}}(z)$ the model flux of the best-fitting template at that redshift.
Taking the logarithm of both sides, the log-likelihood can be separated into two terms:
\begin{equation}
    \ln{\hat{\mathcal{L}}}(z) = -\frac{1}{2}\sum_{i}^{N} \ln{2\pi\sigma_{i}(z)} - \frac{1}{2} \sum_{i}^{N} \left ( \frac{(f_{i,\text{obs}} - \hat{f}_{i,\text{t}})^{2}}{\sigma_{i}(z)^2} \right ),
\end{equation}
where the right-hand term can be rewritten as
\begin{equation}\label{eq:terr_corr_chi}
     \ln{\hat{\mathcal{L}}}(z) = -\frac{1}{2}\sum_{i}^{N} \ln{2\pi\sigma_{i}(z)} - \frac{\hat{\chi}^{2}(z)}{2},
\end{equation}
with $\hat{\chi}^{2}(z)$ the best-fit $\chi^{2}$ at each redshift step of the template fitting (as calculated and stored by \textsc{Eazy}). When deriving the full photo-$z$ posterior, we therefore calculate the first term of the right-hand side of Eq.~\ref{eq:terr_corr_chi} for every source after the fact and incorporate alongside magnitude priors.

For all three fields, we use an IRAC 4.5$\mu$m prior derived from the Bo\"{o}tes spectroscopic sample following the method outlined in Section~5.1.1 of \citetalias{Duncan:2017wu}.
The use of the magnitude prior for the photo-$z$ estimates was found to improve overall statistical performance at $z < 2$ (reduced scatter), the key area of interest for deep fields. 

\subsection{Gaussian process redshift estimates}\label{sec:empirical}
Following previous iterations of our hybrid redshift methodology, we incorporate machine learning estimates derived using the Gaussian process redshift code \textsc{GPz} \citep{2016MNRAS.455.2387A,2016MNRAS.462..726A}.
As in \citetalias{Duncan:2017ul}, when training the \textsc{GPz} classifiers, we employ the magnitude and colour-based weighting scheme \citep[based on the method presented in ][]{Lima:2008eu} to benefit from \textsc{GPz}'s cost-sensitive learning features.
We train \textsc{GPz} using 25 basis functions and allowing variable covariances for each basis function \citep[i.e. the `GPVC' of][]{2016MNRAS.455.2387A}.
Finally, we also follow the practices outlined in Section 6.2 of \citet{2016MNRAS.455.2387A} and allow pre-processing of the input data to normalise or de-correlate the features (also known as `sphering' or `whitening').

A key change to our method from previous implementations is the choice of input magnitudes. 
While the use of standard logarithmic magnitudes has produced good photo-$z$ results in our previous efforts \citepalias{2019A&A...622A...3D}, they are however not suited to datasets containing large numbers of non-detections (i.e. flux measurements with $\text{S/N} < 2$) or low S/N measurements.
In multi-wavelength forced photometry catalogues such as those employed here where depth varies significantly between filters (Fig.~\ref{fig:photom_depths_violin}), non-detections in individual bands are inevitable and therefore a number of sources may have zero or negative flux measurements and hence are undefined in standard logarithmic magnitudes.
In many cases, most notably high-redshift galaxies, those flux limits provide valuable colour (and hence redshift) information.

We therefore make use of \emph{asinh} magnitudes \citep[][sometimes referred to as `luptitudes']{1999AJ....118.1406L} for \textsc{GPz} analysis, since they are able to incorporate zero or negative flux measurements and therefore allow us to train using all available measurements.\footnote{We reiterate that magnitudes are used only for \textsc{GPz} estimates. Template fitting is performed using flux measurements.}
For a given flux, $f$ (with a flux zeropoint $f_{0} = 3631~\rm{Jy}$), we define the \emph{asinh} magnitudes as
\begin{equation}
	 m = \frac{-2.5}{\log(10)} \times \sinh^{-1}\left ( \frac{f/f_{0}}{2b} \right ) + \log(b),
\end{equation}
where the softening parameter, $b$, for each band is derived from the median flux uncertainty on $\approx 5 \sigma$ sources across the field (specifically, $4.95 -  5.05\sigma$).\footnote{The corresponding magnitude uncertainties are then given by $\sigma_{m} =  \frac{-2.5}{\log(10)} \times \frac{(\sigma_{f}/|f|)}{ \sqrt{\left(1 + (2b / (f/f_{0}))^{2}\right )}}$.}
For high S/N measurements, the choice of flux zeropoint ensures that the \emph{asinh} magnitudes are equal to traditional AB magnitudes.
 As demonstrated in \citet{2019MNRAS.489..820B}, photo-$z$ estimates from \emph{asinh} magnitudes are typically not sensitive to the details of the softening parameter used if the data are of similar depth across the field.
For individual photometric bands within the deep fields optical datasets, this assumption is valid due to the relatively homogeneous depth within our datasets (the heterogeneity within the fields is largely a result of different spatial coverage of different bands).
 
Due to the differing range and sources (i.e. telescopes or surveys) of optical to mid-IR photometric imaging across the three fields, \textsc{GPz} must be trained separately for each of the corresponding spectroscopic datasets.
The primary \textsc{GPz} classifier for each field is trained on optical sources that do not satisfy any of the AGN selection criteria - corresponding to the significant majority of both the training sample and photometric catalog.
 In Bo\"{o}tes, we also use additional \textsc{GPz} classifiers trained on the IR, optical and X-ray selected AGN subsets.
 However, due to the lack of comparable quality X-ray imaging, in the EN1 and LH fields we use additional IR and optical AGN \textsc{GPz} estimates only.
 As illustrated in \citetalias{Duncan:2017wu}, sources can satisfy multiple AGN selection criteria.
 The resulting spectroscopic training samples for the different AGN selections are as follows: the IR, Optical and X-ray AGN sample sizes are 1936, 1750 and 1307 (2714 in total) for Bo\"{o}tes, while for EN1 and LH the IR and Optical AGN sample sizes are 346 and 563 (593), and 579 and 1152 (1182), respectively.
  For all training, each input sample was split at random into training, validation and test samples consisting of 80\% (training), 10\% (validation), and 10\% (test) of the full sample, respectively.
  
 The photometric bands used for \textsc{GPz} training for each subset and field are listed in Table~\ref{tab:gpz_bands}.
 The exact choice of bands used was based on balancing the maximum number of bands (and hence information) with the fraction of each field covered by the respective bands.
 We note that the relative merit of including specific bands in the \textsc{GPz} estimates varies between fields due to the variation in depth and coverage of different bands (for example, less improvement was gained by including IRAC 5.8 and 8.0$\mu$m photometry for the galaxies sample in EN1 - see Table~\ref{tab:gpz_bands}).

\begin{table*}
\centering
\caption{Photometric bands used for \textsc{GPz} estimates in each field. See Kondapally et al. (2020) for details on depths and spatial coverage. For EN1 and LH, IRAC photometry is taken from SWIRE only.}\label{tab:gpz_bands}
\begin{tabular}{C{3.0cm}|C{4.5cm} C{4.5cm} C{4.5cm}}
\hline
Subset  &  Bo\"{o}tes & EN1 & LH  \\
   \hline
   \hline
Galaxies & $u$, $B_{\text{w}}$, $R$, $I$, $J$, $H$, $K_{\text{s}}$, \newline 3.6$\mu$m, 4.5$\mu$m, 5.8$\mu$m, 8.0$\mu$m &  $u$, $g_{\text{PS}}$, $r_{\text{PS}}$, $i_{\text{PS}}$, $z_{\text{PS}}$, $y_{\text{PS}}$, $J$, $K$, 
             3.6$\mu$m, 4.5$\mu$m & $u_{}$, $g_{}$, $r_{}$, $z_{}$, 3.6$\mu$m, 4.5$\mu$m, 5.8$\mu$m, 8.0$\mu$m  \\
AGN \newline (IR, Opt, X-ray$^\star$)  & $u$, $B_{\text{w}}$, $R$, $I$, $J$, $H$, $K_{\text{s}}$, \newline 3.6$\mu$m, 4.5$\mu$m, 5.8$\mu$m, 8.0$\mu$m & $u$, $g_{\text{PS}}$, $r_{\text{PS}}$, $i_{\text{PS}}$, $z_{\text{PS}}$, $y_{\text{PS}}$, $J$, $K$, 3.6$\mu$m, 4.5$\mu$m, 5.8$\mu$m, 8.0$\mu$m & $u_{}$, $g_{}$, $r_{}$, $z_{}$, 3.6$\mu$m, 4.5$\mu$m, 5.8$\mu$m, 8.0$\mu$m\\
\hline
  \end{tabular}
\end{table*}

\subsection{Calibration of photo-$z$ uncertainty}\label{sec:pz_accuracy_method}
As outlined in Section~\ref{sec:photoz-modelerrors}, a key goal of the modifications to our template fitting method is to incorporate model uncertainties.
Nevertheless, additional calibration of the uncertainties on the resulting photo-$z$ is still necessary.
To quantify the over- or under-confidence of our photo-$z$ estimates, we follow the method outlined in Section~3.3.1 of \citetalias{Duncan:2017ul} \citep[and originally proposed by ][]{2016MNRAS.457.4005W} and calculate the distribution of threshold credible intervals, $c$, where the spectroscopic redshift intersects the redshift posterior.
For perfectly accurate estimates of the uncertainties, the cumulative distribution of credible intervals, $\hat{F}(c)$, should follow a straight 1:1 relation, i.e. a quantile-quantile (or $Q-Q$) plot.
Curves that fall below this 1:1 relation indicate that there is overconfidence in the photo-$z$ errors (i.e. the $P(z)$s are too sharp) while curves that fall above indicate under-confidence.

Following \citetalias{Duncan:2017wu}, we scale the uncertainties on the template fitting estimate for a source $i$, such that 
\begin{equation}\label{eq:smoothing_1}
	P(z)_{\textup{new}, i} \propto P(z)_{\textup{old}, i}^{1/\alpha(m_{i})} \times P(z|m_{i}),
\end{equation}
\noindent where $\alpha(m)$ is a magnitude-dependent function following:
\begin{equation}\label{eq:smoothing_2}
	\alpha(m) = \begin{cases}
	 \alpha_{\eta} & m \leq m_{\eta}\\
	 \alpha_{\eta} + \kappa \times(m-m_{\eta}) & m > m_{\eta},
	\end{cases}
\end{equation}
\noindent with $\alpha(m)$ being constant ($\alpha_{\eta}$) below some characteristic apparent magnitude, $m_{\eta}$, and following a simple linear relation above this magnitude \citep{2009ApJ...690.1236I}.
For Bo\"{o}tes, EN1, and LH we use the $I$, $r_{\text{PS}}$ and $r$-band optical \emph{asinh} magnitudes, respectively for calculating the magnitude dependence of the error scaling and assume a characteristic magnitude of $m_{\eta} = 16$.
The parameters $\alpha_{\eta}$ and $\kappa$ are then fit using the \textsc{emcee} Markov chain Monte Carlo fitting tool \citep[MCMC;][]{2013PASP..125..306F} to minimise the Euclidean distance between the measured and ideal $\hat{F}(c)$ distributions.

When calibrating the uncertainties produced by \textsc{GPz}, we calculate the threshold credible interval following 
$c_{i} = \textup{erf} \left (| z_{i,\textup{spec}} - z_{i,\textup{phot}} | / \sqrt{2} \sigma_{i} \right)$ \citepalias{2019A&A...622A...3D} and scale the uncertainties as a function of magnitude in the same manner as for the template estimates:
\begin{equation}\label{eq:gpz_err_scale}
	\sigma_{\textup{new},i} = \sigma_{\textup{old},i} \times \alpha(m_{i}),
\end{equation}
\noindent where $\alpha(m_{i})$ follows the same functional form as Eq.~\ref{eq:smoothing_2}.
As above, the parameters for $\alpha(m_{i})$ are optimised through MCMC minimisation of the difference between the measured and ideal $\hat{F}(c)$ distributions.

As highlighted in Fig.~\ref{fig:specz_magdist}, the available spectroscopic sample is biased towards bright optical magnitudes.
Even though our photo-$z$ uncertainties are calibrated as a function of magnitude, the biased sample could result in improved photo-$z$ uncertainty accuracy for the brightest sources at the expense of the faint population. 
In order to prevent the optimisation of the photo-$z$ uncertainties being dominated by the most populous magnitude ranges, we calibrate the uncertainties using a magnitude balanced subset of the total spec-$z$ population following the approach presented in \citetalias{2019A&A...622A...3D}.
For both the AGN and galaxy samples separately, a subset of each training sample is created by randomly selecting up to 750 sources in magnitude bins (of width $=1$ mag) over the range covered by the spectroscopic redshift subset.
Calibration of the uncertainties is then done on two-thirds of this subsample, with the other one-third retained for testing.
We note that this balancing of the spec-$z$ sample is most important for the Bo\"{o}tes field where the larger samples of bright magnitude selected spec-$z$s could significantly bias the optimisation.

\subsection{Hierarchical Bayesian combination}\label{sec:hbcombination}
To produce the final consensus redshift prediction for a given source, we use the HB combination method outlined by \citetalias{Duncan:2017wu} \citep[based on the method presented in][]{Dahlen:2013eu} and subsequently extended to hybrid \textsc{GPz} + template estimates in \citetalias{Duncan:2017ul}.
In summary, the HB combination produces a consensus redshift prediction, $P(z)$ from a set of $n$ individual predictions while marginalising over the probability that any individual $P(z)$ is incorrect.
Hyper-parameters for the fraction of measurements that are bad, $f_{\text{bad}}$, and the relative covariance between the different estimates $\beta$, are optimised using training data to ensure that the posterior redshift distributions more accurately represent the redshift uncertainties.

As in \citetalias{2019A&A...622A...3D}, \textsc{GPz} estimates are evaluated on the same redshift grid as used during the template fitting procedure.
If a source does not have a photo-$z$ estimate for a given \textsc{GPz} estimator (either through not satisfying the selection criteria for a given subset or lack of observations in a required band) it is assumed to have a flat redshift posterior for that specific estimator.
\textsc{GPz} therefore contributes no additional information to the consensus HB estimates for these sources. 

For the application in this work, we assume $0 \leq f_{\text{bad}} \leq 0.05$ and $0 \leq f_{\text{bad}} \leq 0.2$ for the galaxy and AGN subsets, respectively, and a flat prior on the redshift distribution for `bad' estimates.
However, one further improvement to our method in this work is that we optimise the hyper-parameter $\beta$ (i.e. the degree of covariance between different photo-$z$ estimates) as a function of magnitude.
This change is necessary to ensure that the resulting consensus photo-$z$ posteriors provided accurate uncertainties across all magnitude ranges. 
Assuming a constant $\beta$ for all magnitudes results in underestimates of the uncertainties for the brightest and faintest sources.


In the following section, we present analysis of the consensus photo-$z$ estimates for each field.


\section{Photometric redshift properties }\label{sec:results}
\begin{figure*}[h!]
\centering
 \includegraphics[width=0.9\textwidth]{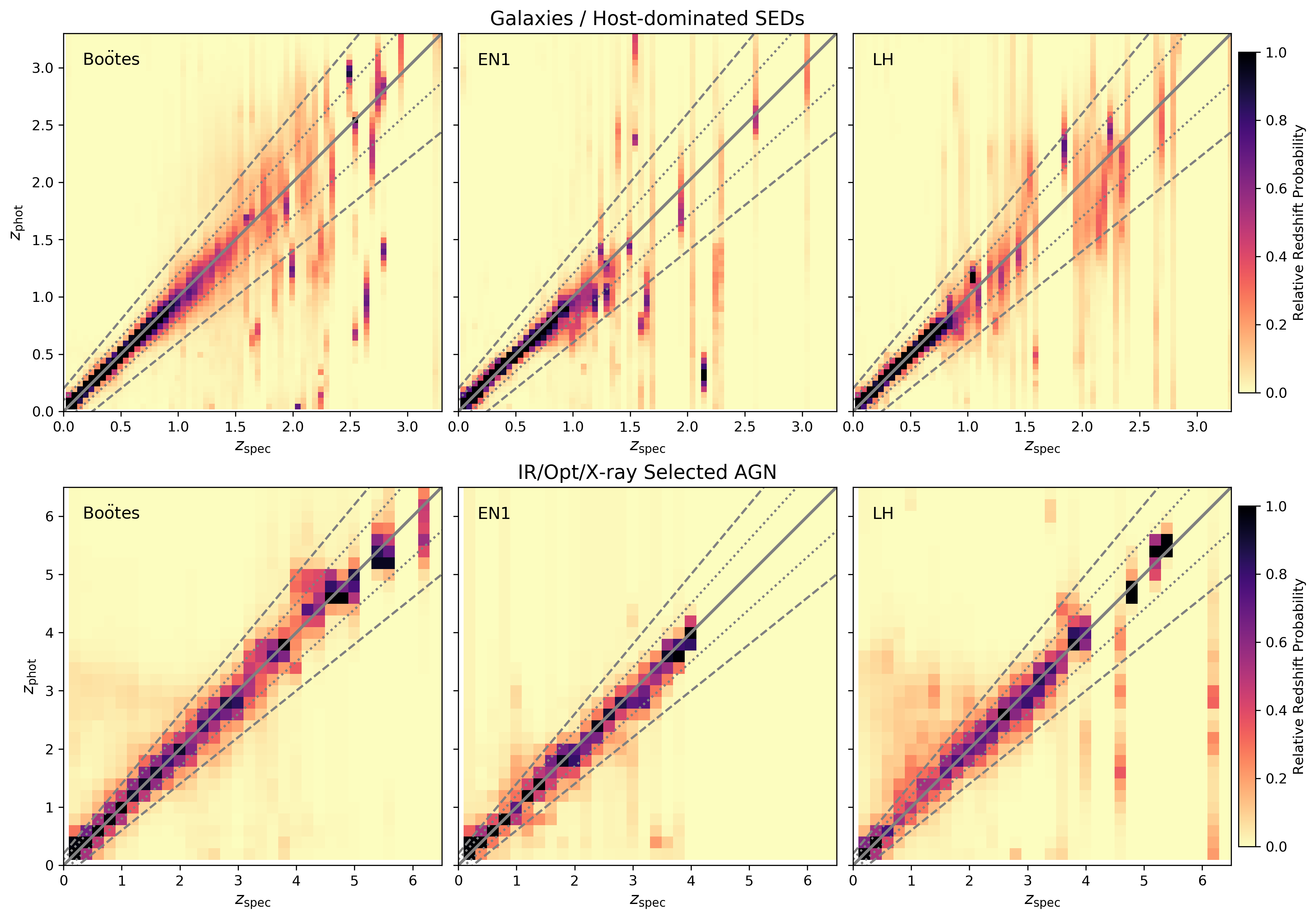}
 \caption{Stacked redshift posterior estimates for the galaxy (or host-dominated) population (top) and for the combined AGN selected population (bottom; IR, X-ray or optically selected) for each of the three fields. The spectroscopic redshift samples available become particularly sparse at $z > 1$ for the EN1 and LH fields. Grey dotted and dashed line correspond to $\pm 0.1\times(1+z_{\textup{spec}})$ and $\pm 0.2\times(1+z_{\textup{spec}})$, respectively. The colour scale is illustrative and represents the summed $P(z)$ (normalised such that $\int{P(z)} \text{d}z = 1$) for each $z_{\textup{spec}}$ bin, more precise photo-$z$ therefore result in darker peaks. }
 \label{fig:specz_photz_pz}
\end{figure*}
In this work we have estimated photo-$z$s for all sources within the optical catalogues for which estimates could be made, including parts of the fields with very limited photometric coverage (including regions outside the area where radio to optical cross-identifications have been performed in \citetalias{Kondapally2020}).
However, due to the heterogeneous photometry coverage within the deep fields, the resulting photo-$z$ quality will also vary across the field.
In the following analysis we examine the photo-$z$ quality within the core regions of each field where photometric data is relatively homogeneous.
These areas broadly conform to those where \textsc{GPz} estimates are available and where radio source cross-matching has been performed for the LoTSS Deep Field data (Paper III).
Specifically, we apply the following cuts in each field based on the flags in the input photometric catalogues of \citetalias{Kondapally2020}:
In Bo\"{o}tes, we require that $\texttt{FLAG\_DEEP} \neq 0$ to remove duplicate or masked sources from the $I$-band detected catalogue sources. 
In EN 1, we require $\texttt{FLAG\_OVERLAP} \geq 6$ to restrict the analysis to the region that includes both NIR and \emph{Spitzer} mid-IR photometric coverage. 
Due to the smaller coverage of the NIR data relative to the optical and mid-IR coverage, in LH we do not require NIR data when analysing the photo-$z$ in this field - restricting the analysis to regions with SpARCS optical coverage ($\texttt{FLAG\_OVERLAP} \geq 6$) which overlaps with the \emph{Spitzer} imaging. 

Additionally, in all fields we require $\texttt{FLAG\_CLEAN} = 1$ to exclude sources within the optical bright star mask.
Compared to the initial combined total of $\approx7.2\times10^{6}$ catalogue sources available across the three fields, these cuts reduce the number of sources with the most robust photo-$z$ estimates to a combined $\approx 5\times10^{6}$ sources.

Fig.~\ref{fig:specz_photz_pz} presents a qualitative illustration of the final consensus redshifts after the error calibration for all input estimates and the tuning of the Bayesian combination hyper-parameters.
We show the stacked redshift posteriors as a function of spectroscopic redshift for each field, with the spectroscopic samples separated into the galaxy or host-dominated (top row) and AGN subsets (bottom row).
From Fig.~\ref{fig:specz_photz_pz} we can see that the photo-$z$ quality for galaxies at $z < 1$ is excellent for all three fields. 
At higher redshifts, the spectroscopic samples become much more limited (especially outside of Bo\"{o}tes).
However, from the limited number of spec-$z$s available in this regime, it is clear that the performance deteriorates between $1 \lesssim z \lesssim 1.5$ for the host-dominated population, and gets substantially worse beyond $z\sim1.5$.
Examining the AGN population posteriors, we find that photo-$z$ performance looks generally good out to $z > 3$ in all fields - albeit with significantly larger scatter than the host-dominated population. 
Few spectroscopically confirmed sources at $z > 4$ exist in the EN1 and LH but we find that our photo-$z$ estimates in Bo\"{o}tes provide accurate estimates out to $z > 6$ \citep[including the confirmed radio-loud quasar at $z=6.1$][]{2006ApJ...652..157M}.
Given that hybrid photo-$z$ estimates in this regime are found to be dominated by the template estimates \citepalias[see][]{Duncan:2017ul}, we would expect the performance in the EN1 and LH datasets to therefore be comparable in quality (or potentially better in the case of EN1).
Nevertheless, we caution against the use of the photo-$z$s above $z\sim4$ without careful analysis of the individual estimates.

\subsection{Overall photo-$z$ statistics}
The $z_{\textup{spec}}$ versus $z_{\textup{phot}}$ distribution shown in Fig.~\ref{fig:specz_photz_pz} allows us to qualitatively assess the photo-$z$ performance. 
However, a more quantitative analysis is required to enable comparison between fields and for defining appropriate selection criteria for future science exploitation.
When calculating photo-$z$ statistics, we use the median of the primary redshift peak ($z_{1,\textup{median}}$) following \citetalias{2019A&A...622A...3D}: see Section~4.1 of that paper for details on how this value is defined.
For our measure of robust scatter, we then use the normalised median absolute deviation,  $\sigma_{\textup{NMAD}}$, defined as:
\begin{equation}\label{eq:nmad}
\sigma_{\textup{NMAD}} =1.48 \times \text{median} ( \left | \Delta z \right | / (1+z_{\textup{spec}})),
\end{equation}
where $\Delta z = z_{1, \textup{median}} - z_{\textup{spec}}$.
Similarly, we define the OLF as sources where 
\begin{equation}\label{eq:olf}
	\left | \Delta z \right | / (1+z_{\text{spec}}) > 0.15,
\end{equation}
as is common in the literature \citep[e.g. ][]{Dahlen:2013eu}.

\begin{table}
\caption{Photo-$z$ quality statistics for the galaxy and AGN spectroscopic redshift samples in each deep field. The samples (of size $N$) are cut to have $z_{\textup{phot}} < 1.5$ for the galaxy subset and $z_{\textup{phot}} < 4$ for the AGN subset to allow a more direct comparison between fields. The robust scatter, $\sigma_{\textup{NMAD}}$, and $\textup{OLF}$, are defined in Equations~\ref{eq:nmad} and \ref{eq:olf}, respectively.}
\centering
\begin{tabular}{lccc}
\hline
\multicolumn{4}{c}{Galaxies / Host-dominated} \\
\hline
 & $N$ & $\sigma_{\textup{NMAD}}$ & OLF \\
 \hline
Bo\"{o}tes & 15200 & 0.016 & 0.018 \\
ELAIS-N1 & 2570 & 0.02 & 0.016 \\
Lockman Hole & 2743 & 0.017 & 0.015 \\
\hline
\multicolumn{4}{c}{AGN}\\
\hline
 & $N$ & $\sigma_{\textup{NMAD}}$ & OLF \\
\hline
Bo\"{o}tes & 2215 & 0.07 & 0.177 \\
ELAIS-N1 & 442 & 0.064 & 0.224 \\
Lockman Hole & 701 & 0.077 & 0.223 \\
\hline
\end{tabular}\label{tab:photz_stats}
\end{table}
\begin{figure*}[]
\centering
 \includegraphics[width=0.8\textwidth]{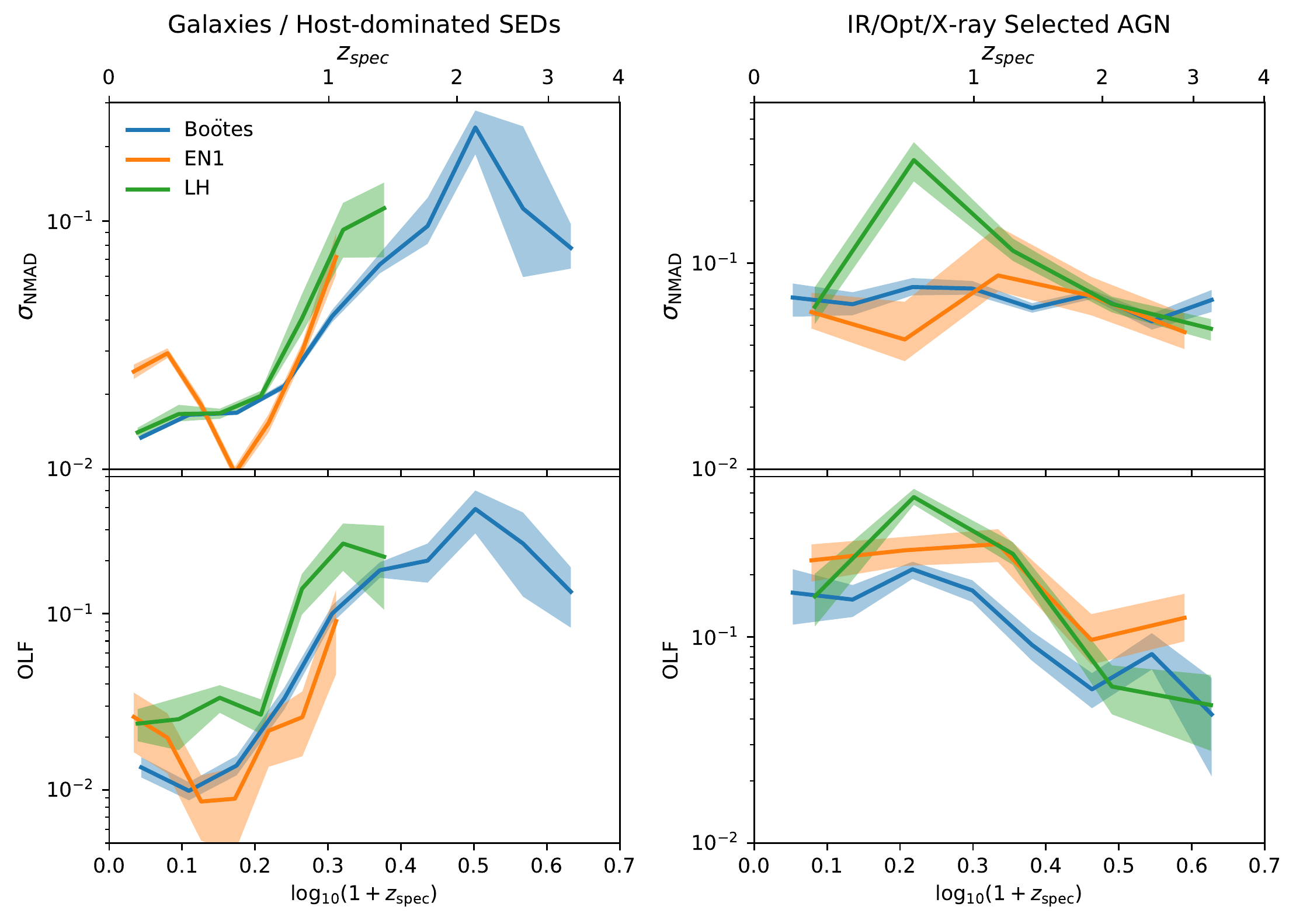}
 \caption{Robust scatter ($\sigma_{\textup{NMAD}}$; upper panels) and $\textup{OLF}$ (lower panels) for the consensus photo-$z$ estimates ($z_{1,\textup{median}}$) as a function of spectroscopic redshift. Shaded regions illustrate the statistical uncertainties on the respective metrics derived from bootstrap resampling.}
 \label{fig:stats_z}
\end{figure*}
\begin{figure*}[]
\centering
 \includegraphics[width=0.8\textwidth]{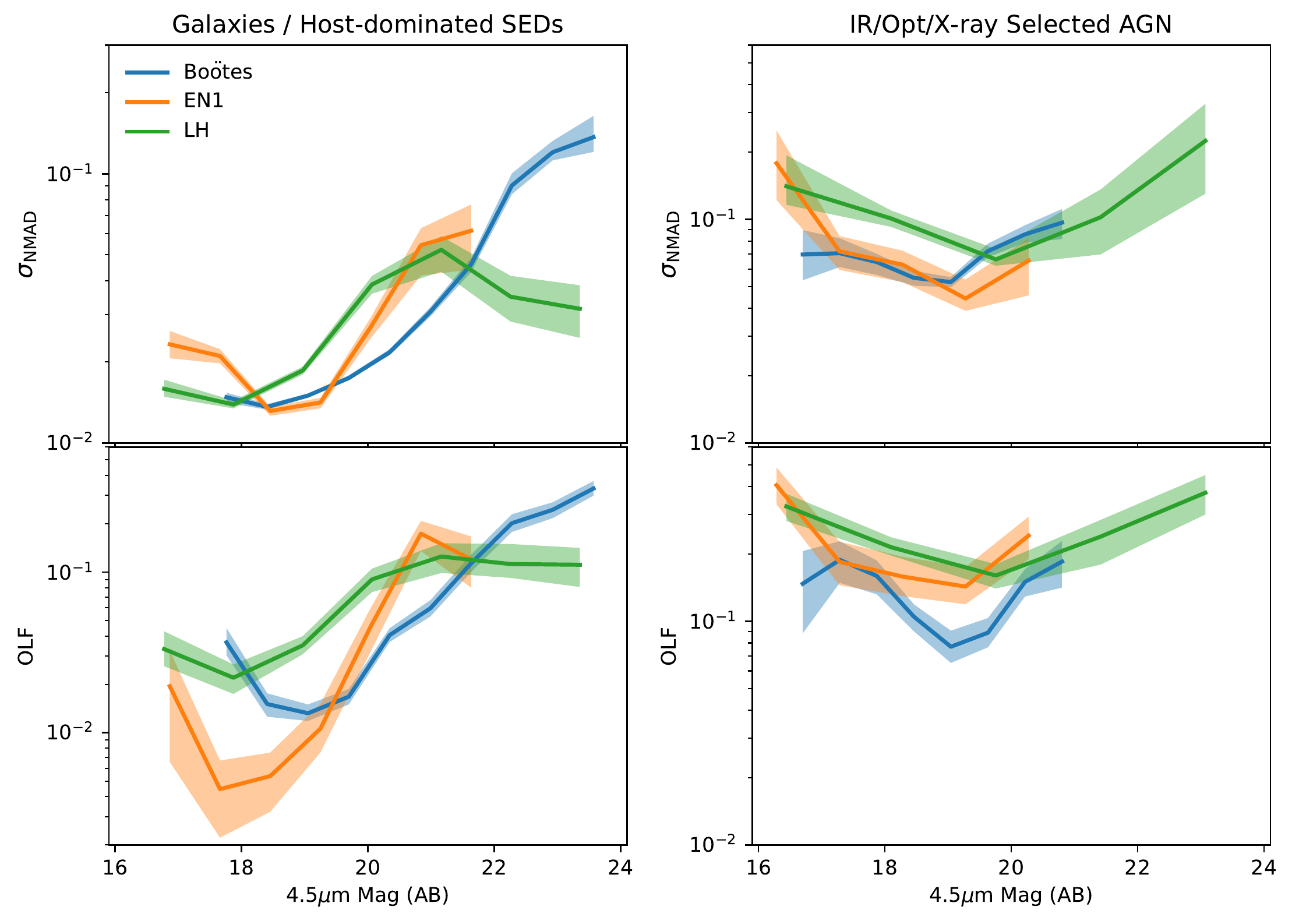}
 \caption{Robust scatter ($\sigma_{\textup{NMAD}}$; upper panels) and $\textup{OLF}$ (lower panels) for the consensus photo-$z$ estimates ($z_{1,\textup{median}}$) as a function of 4.5\,$\mu$m magnitude. Shaded regions illustrate the statistical uncertainties on the respective metrics derived from bootstrap resampling. The magnitude range plotted for each field is representative of the magnitude distribution of the spectroscopic sample and is not indicative of the relative photometric depths (see Fig.~\ref{fig:photom_depths_violin}).}
 \label{fig:stats_mag}
\end{figure*}

In Table~\ref{tab:photz_stats} we present the overall $\sigma_{\textup{NMAD}}$ and OLF statistics derived for the galaxy and identified AGN subsets.
To ensure a fair comparison across the three fields, for the galaxy subset we limit the statistics to those sources with $z_{1,\textup{median}} < 1.5$ while for the AGN subsample we limit to $z_{1,\textup{median}} < 4.0$.
We note that these cuts are defined using $z_{1,\textup{median}}$ (as opposed to $z_{\textup{spec}}$) to better represent the performance expected for science samples that will be defined based on their photo-$z$ alone.

We find that all three fields exhibit very similar $\sigma_{\textup{NMAD}}$ and OLF metrics for both subsets, with less than 2\% scatter and OLF for galaxies or host-dominated sources.
As already seen in Figure~\ref{fig:specz_photz_pz} the photo-$z$ quality for the AGN subset is significantly worse, with OLF reaching $\approx20\%$ or higher.
Since the available spectroscopic redshifts are biased to typically brighter galaxies and lower redshifts, these statistics are not fully representative of the true photo-$z$ accuracy and reliability for the average galaxy population.

In Fig.~\ref{fig:stats_z} and \ref{fig:stats_mag} we present the resulting robust scatter and OLF as a function of redshift and magnitude, respectively.
The resulting statistics provide a quantitative confirmation of the photo-$z$ quality visible in Fig.~\ref{fig:specz_photz_pz}, where we find that our photo-$z$ estimates have excellent scatter $<0.04\times(1+z)$ and OLF $\sim5\%$ for the galaxy/host-dominated population at $z < 1$ across all three fields. 
For all three fields, there is a deterioration in the photo-$z$ scatter and OLF with increasing redshift.
However, in LH and EN1, not enough spectroscopic redshifts are available above $z\sim1$ redshift to assess the quality beyond this point.
In Bo\"{o}tes, where spec-$z$s are available in this regime, we find that the trend in the statistics verifies the visual interpretation of Figure~\ref{fig:specz_photz_pz} - whereby the measured scatter deteriorates above $z\sim1$ before becoming substantially worse at $z > 1.5$. 

In contrast to the strong correlation with redshift observed for the galaxy subset, the measured $\sigma_{\textup{NMAD}}$ for the identified optical, IR and X-ray AGN samples exhibits negligible evolution with redshift.
We also find that the OLF appears to decline as a function of redshift.
This trend is likely driven by biases within the spectroscopic sample (with the $z > 2$ training and test samples are typically dominated by optically bright QSOs) and the presence of the strong Lyman break feature redshifting into the $u$-band at $z\sim3$, leading to improved template fitting estimates \citepalias[see also][for further discussion]{Duncan:2017ul}.

We find that the measured OLF for known AGN at a given redshift or 4.5$\mu$m magnitude in EN1 and LH is higher than for Bo\"{o}tes. 
We note that this difference is likely driven by the demographics of the respective spectroscopic redshift samples available.
The EN1 and LH spec-$z$ samples are dominated by QSOs while the Bo\"{o}tes sample includes a large number of sources explicitly selected on X-ray or IR AGN criteria.
In \citetalias{2019A&A...622A...3D}, photo-$z$ performance for AGN selected as optical QSOs was measured to be significantly worse than for those selected based on the other AGN criteria.

\subsubsection{Comparison with existing literature photo-$z$ estimates}\label{sec:photz_lit_comp}

In \citetalias{Duncan:2017ul} we compared the quality of photo-$z$ produced by our hybrid methodology with existing estimates in the literature, finding that our results produced better statistics than redshifts presented by \citet[][]{Brodwin:2006dp}, with the substantial benefits of the accurate redshift posteriors.
Thanks to the improved template fitting methodology in this study, the results presented in this data release represent a further improvement in photo-$z$ quality available for this field.
Specifically, when calculating photo-$z$ quality statistics for the same subset of spectroscopic sources as used in Table~\ref{tab:photz_stats}, the photo-$z$ presented in \citetalias{Duncan:2017ul} have $\sigma_{\textup{NMAD}} = 0.032$ and OLF$ = 0.028$ for galaxies and $\sigma_{\textup{NMAD}} = 0.12$ and OLF$ = 0.31$ for the known AGN.
The results presented in this paper therefore represent up to a factor of $\sim2$ improvement in redshift quality averaged over the available spectroscopic population (we note that the degree of improvement in parameter space not probed by the existing spec-$z$ sample could differ substantially from this, both positively and negatively).

In the EN1 and LH fields, there are a number of recently published photo-$z$ catalogues to which we can directly compare our results.
As part of the Hyper Suprime-Cam Subaru Strategic Program Second Public Data-release \citep[HSC PDR2;][]{2019PASJ...71..114A}, photo-$z$ estimates derived from the HSC optical photometry (\textit{g,r,i,z,y}) are available for the EN1 field for much of the area covered by the LoTSS Deep Field optical catalogues.
The HSC PDR2 photo-$z$ estimates presented by \citet[][see also \citeauthor{2018PASJ...70S...9T}\, \citeyear{2018PASJ...70S...9T}]{2020arXiv200301511N} are based on two different approaches: empirical estimates using the `Direct Empirical Photometric method' \citep[DEmP][]{2014ApJ...792..102H}, and template based estimates following the method presented in \citep[][\textsc{Mizuki}]{2015ApJ...801...20T}.

To directly compare our results with the HSC PDR2 estimates, we cross-match the catalogues with a maximum separation of 1\arcsec and calculate the photo-$z$ quality statistics for all three estimates (this work, HSC DeMP and HSC \textsc{Mizuki}) for the subset of spectroscopic sources with measurements in both our results and at least one of the HSC catalogues.
The resulting statistics, presented in Table~\ref{tab:photz_hsc_stats}, demonstrate that the estimates presented in this work are again comparable to or better than the existing literature.
While the overall statistics for the galaxy/host-dominated AGN population offer only a small improvement on the excellent scatter and OLF of the HSC estimates, our estimates for the known AGN population offer significantly better precision and reliability.
This improvement is even starker when restricting the analysis to the higher redshift AGN population ($z > 1$, for which we provide statistics in Table~\ref{tab:photz_hsc_stats} in parentheses), where our results have lower $\sigma_{\textup{NMAD}}$ and OLF by up to a factor of $\sim3$ compared with those available from HSC PDR2. 

\begin{table}
\caption{Photo-$z$ quality statistics for the galaxy and AGN spectroscopic redshift samples in EN1 in comparison to literature values from the HSC PDR2 \citep{2020arXiv200301511N}. The samples (of size $N$) are cut to have $z_{\textup{phot}} < 1.5$ for the galaxy subset and $z_{\textup{phot}} < 4$ for the AGN subset as in Table~\ref{tab:photz_stats} and are also limited . For the AGN subset, we also show in parentheses the statistics corresponding to limited to sources at $z_{\textup{spec}} > 1$.}
\centering
\begin{tabular}{lccc}
\hline
\multicolumn{4}{c}{Galaxies / Host-dominated} \\
\hline
 & $N$ & $\sigma_{\textup{NMAD}}$ & OLF \\
 \hline
This paper & 2439 & 0.019 & 0.016 \\
HSC DeMP & 2439 & 0.019 & 0.027 \\
HSC Mizuki & 2414 & 0.025 & 0.031 \\
\hline
\multicolumn{4}{c}{AGN ($z > 1$)}\\
\hline
 & $N$ & $\sigma_{\textup{NMAD}}$ & OLF \\
\hline
This paper  & 420 (247) & 0.058 (0.058) & 0.193 (0.158)\\
HSC DeMP & 420 (247) & 0.069 (0.144) & 0.317 (0.445) \\
HSC Mizuki & 318 (164) & 0.09 (0.188) & 0.349 (0.457)\\
\hline
\end{tabular}\label{tab:photz_hsc_stats}
\end{table}

Additionally, \citet{2019MNRAS.483.3168P} present template based photo-$z$ estimates for 18\,deg$^2$ of multi-wavelength optical photometry in the Spitzer Extragalactic Representative Volume Survey \citep[SERVS;][]{2012PASP..124..714M} over five fields, including subsets of the EN1 and LH fields.
 \citet{2019MNRAS.483.3168P} quote robust scatter and outlier statistics calculated using the same definitions as in this paper for their full spectroscopic test sample (and the subset with the best available photometry), finding $\sigma_{\textup{NMAD}} = 0.042\,(0.037)$ and $\text{OLF} =  0.105\,(0.028)$ in the EN1 field and $\sigma_{\textup{NMAD}} = 0.067\,(0.03)$ and $\text{OLF} = 0.205\,(0.048)$ in LH.
As the photo-$z$ estimates in \citet{2019MNRAS.483.3168P} are primarily designed for galaxies (with no AGN templates included in the fitting), we can compare these values with the statistics for the galaxy population presented in Table~\ref{tab:photz_stats}. 
It is clear that the results presented in this work represent a substantial improvement in overall photo-$z$ quality.

Our improved photo-$z$ precision and reliability for the EN1 and LH fields highlights the benefits of not just our photo-$z$ methodology but also the advantages of the full aperture matched photometry across the UV to mid-IR regime provided by \citetalias{Kondapally2020}.
In particular, the availability of NIR and mid-IR photometry is crucial for providing reliable estimates for the AGN population at higher redshifts (a key area of scientific interest for the LoTSS Deep Fields).
Furthermore, the additional colour information provided by forced photometry across all available photometric bands is vital, offering huge improvements in photo-$z$ estimates or SED fitting when compared to the cross-matched catalogues employed by \citet[][see also \citeauthor{2017ApJS..230....9N}~\citeyear{2017ApJS..230....9N}]{2019MNRAS.483.3168P}.

\subsubsection{Photo-$z$ properties for the LOFAR population}
\begin{figure}[h!]
\centering
 \includegraphics[width=1.0\columnwidth]{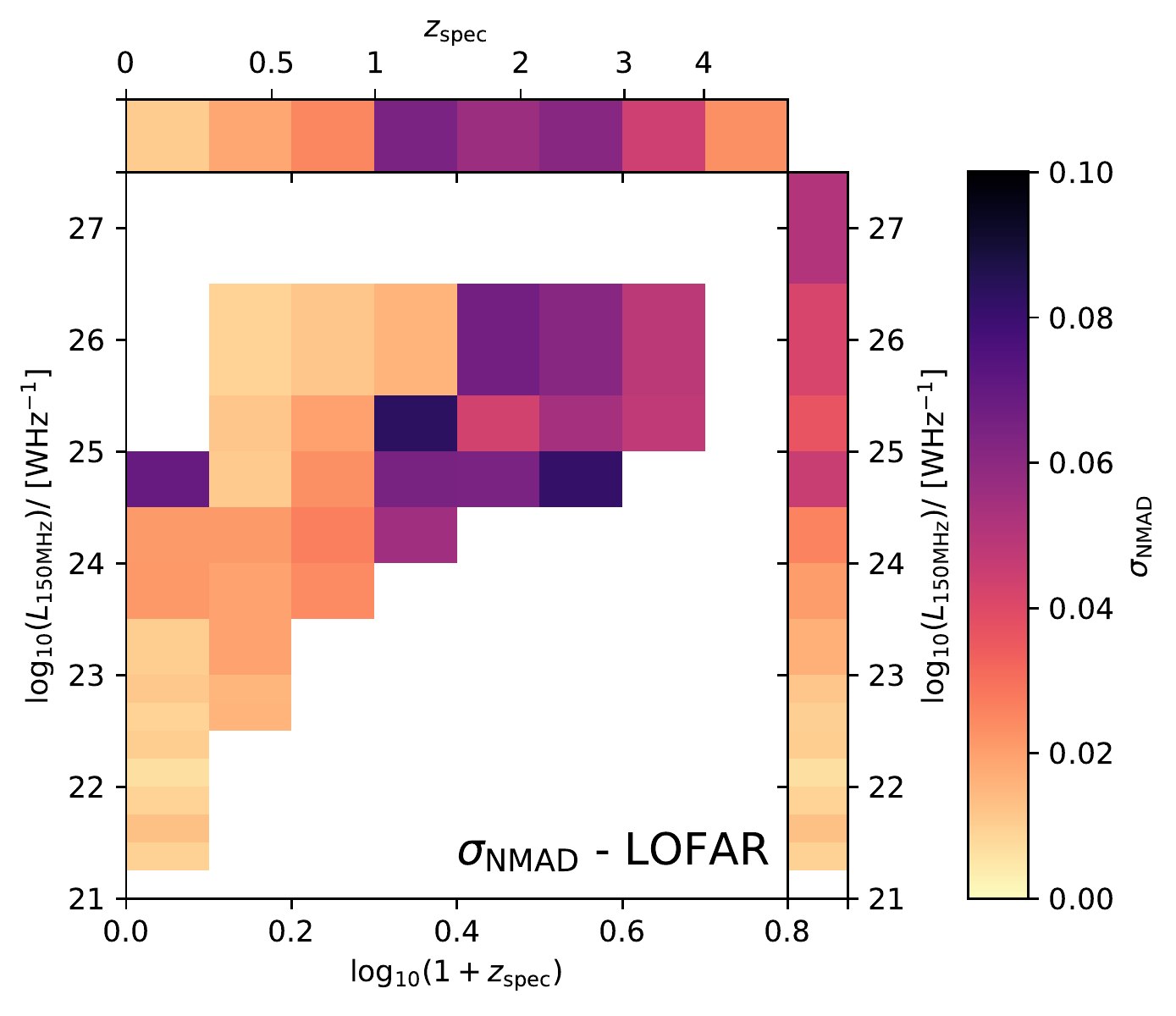}
 \includegraphics[width=1.0\columnwidth]{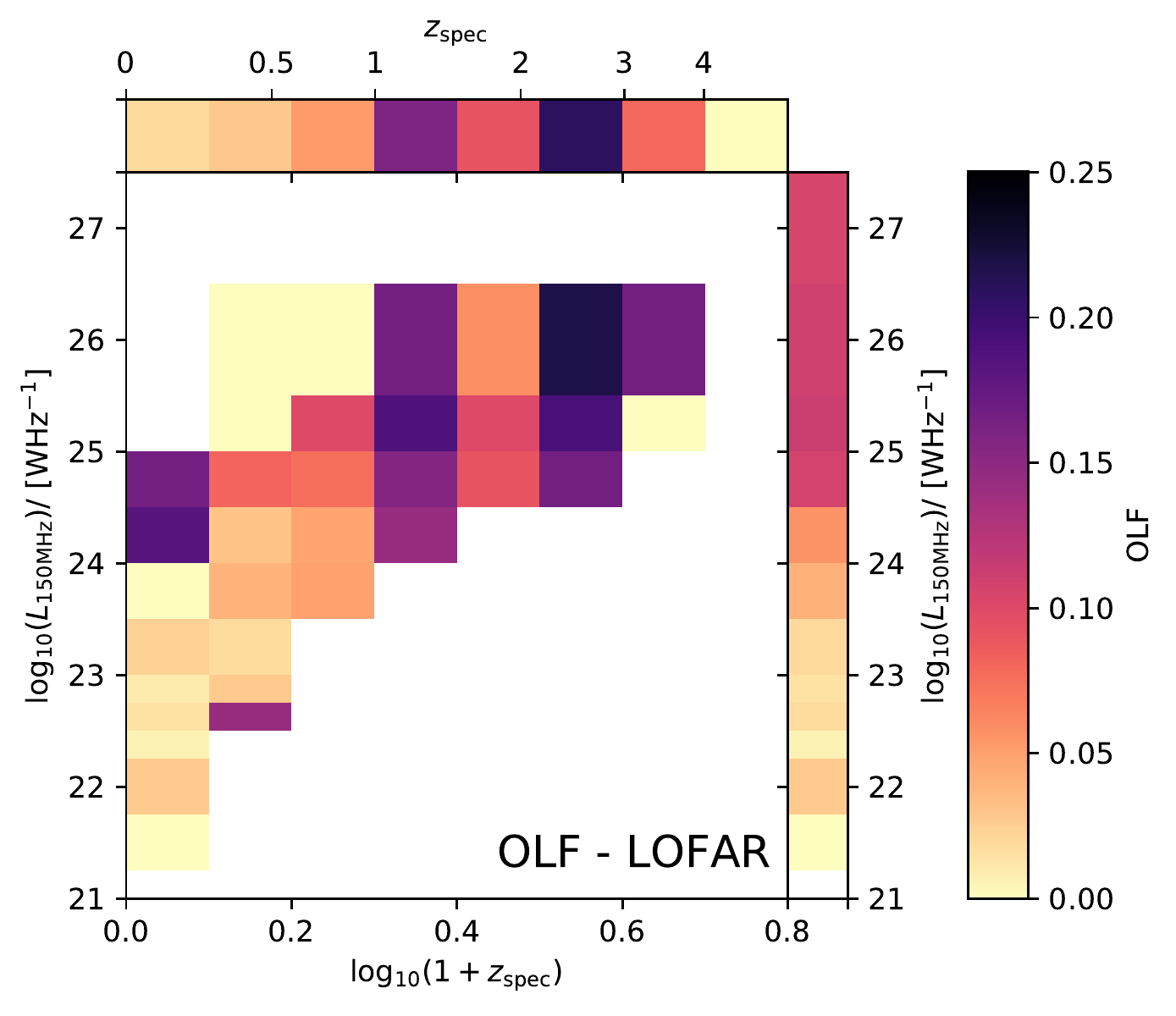}
 \caption{Robust scatter ($\sigma_{\textup{NMAD}}$; top) and $\textup{OLF}$ (bottom) for the consensus photo-$z$ estimate as a function of spectroscopic redshift and 150 MHz radio continuum luminosity in the Bo\"{o}tes field. The top and side panel show the trends averaged over all redshifts and luminosities, respectively. For a cell to be plotted we require a minimum of five galaxies - for some redshifts (or luminosities) we only have the number statistics available to plot the statistics averaged over all luminosities (or redshift).}
 \label{fig:stats_radio}
\end{figure}

Finally, we explore the quality of the consensus photo-$z$ estimates as a function of their radio properties.
Due to the larger spectroscopic sample required to bin in multiple properties, we explore the statistics only in the Bo\"{o}tes field.
However, given the similar performance and redshift or magnitude trends observed across the three fields we would expect any observed trends to hold for all fields.
Figure~\ref{fig:stats_radio} presents the $\sigma_{\textup{NMAD}}$ and $\textup{OLF}$ as a function of both spectroscopic redshift and 150MHz radio luminosity.
When converting from observed flux density to rest-frame radio luminosity, we assume an average spectral slope of $\alpha = -0.7$ for all sources - consistent with the typical slope observed in previous studies \citep{2017MNRAS.469.3468C}

In \citetalias{2019A&A...622A...3D} we observed a clear evolution in the scatter and OLF of the radio source population with $z_{\textup{spec}}$, whereby the photo-$z$ properties of the highest redshift sources are significantly worse than for sources with similar radio luminosity at low redshift.
However, within a given spectroscopic redshift bin, we found no evidence for any significant trend with radio luminosity in either the scatter or OLF.

Although exhibiting a noisier evolution than observed in \citetalias{2019A&A...622A...3D}, our photo-$z$ estimates follow the same overall trend as a function of redshift and radio power.
Specifically, for a fixed redshift we do not observe a strong correlation in $\sigma_{\textup{NMAD}}$ or $\textup{OLF}$ as a function of radio luminosity.
Averaged over all redshifts, the measured scatter and OLF increase with increasing radio power.
However, this trend is clearly driven by the redshift distribution of spectroscopic sources available at a given radio power.

\subsection{Accuracy of the photo-$z$ uncertainties}\label{sec:pz_accuracy_results}
\begin{figure*}[h!]
\centering
 \includegraphics[width=1.0\textwidth]{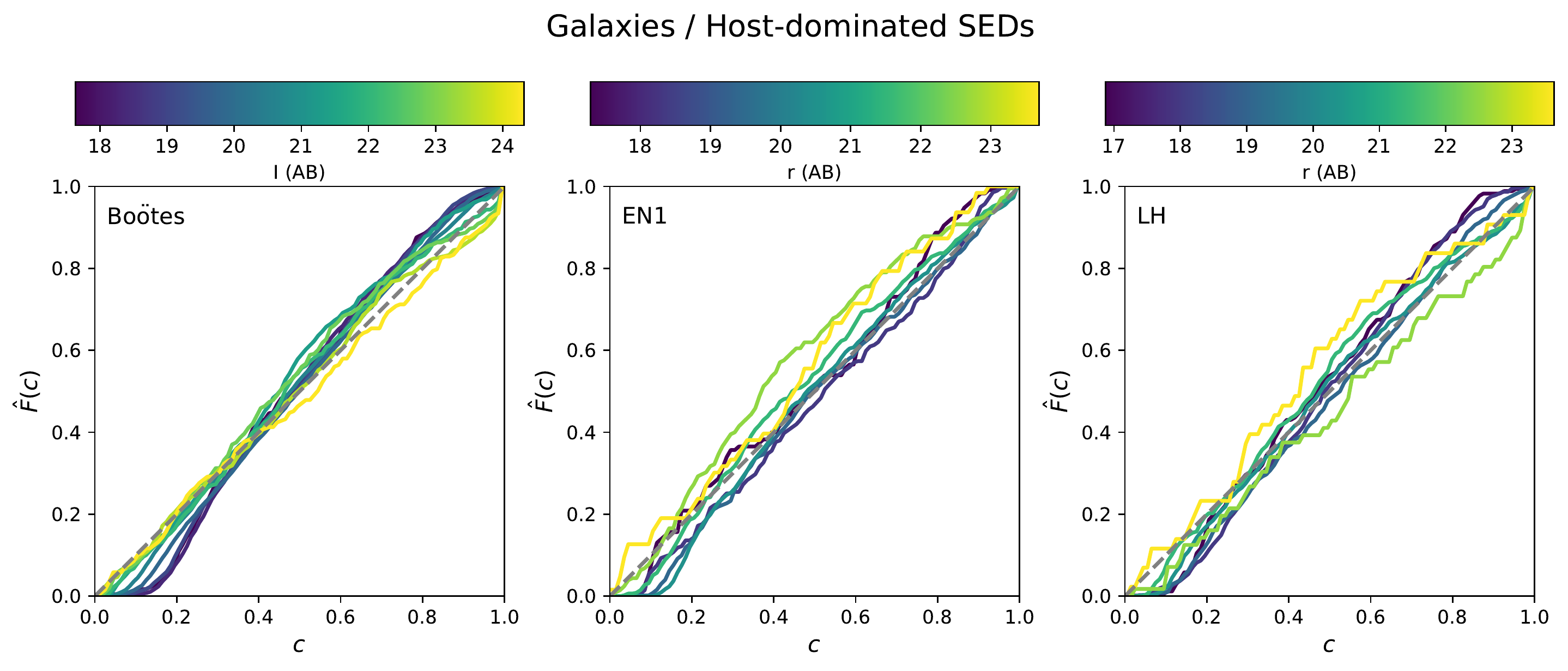}
 \includegraphics[width=1.0\textwidth]{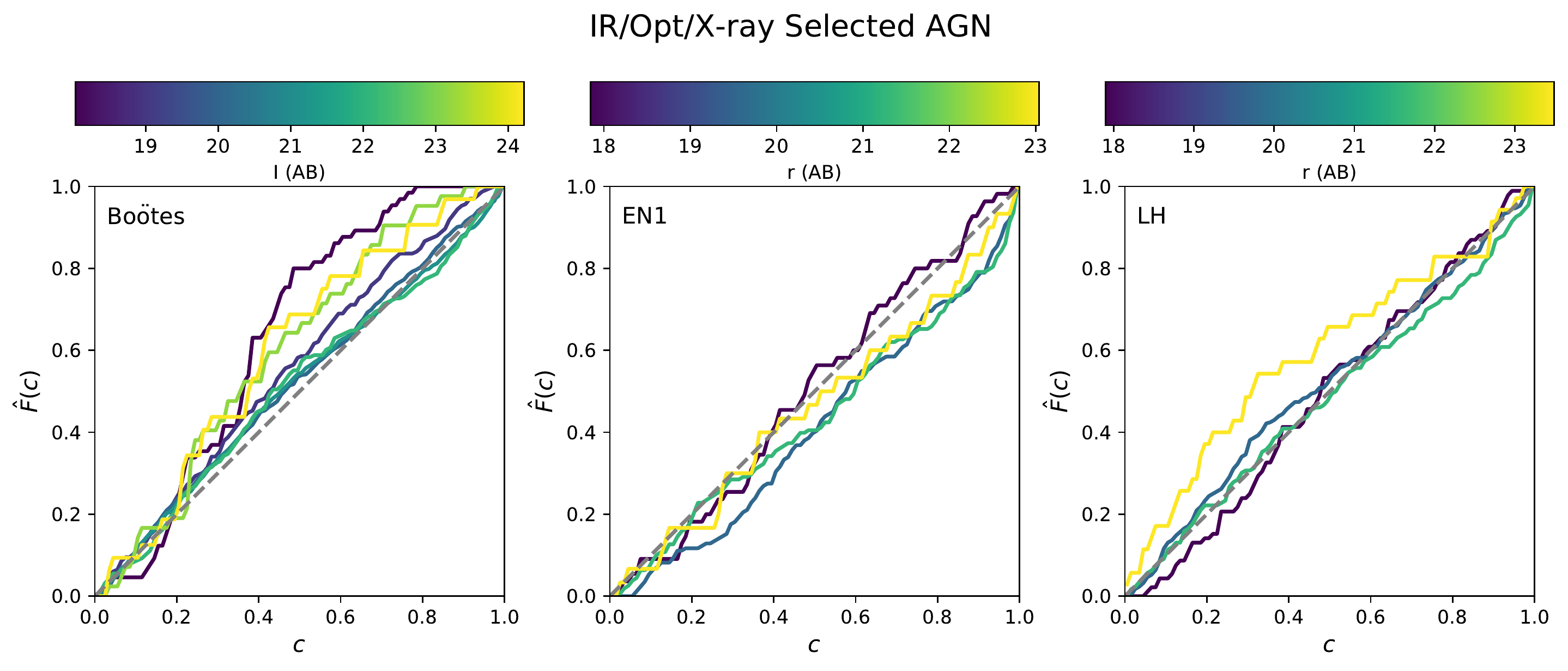}
 \caption{Panels illustrate the cumulative distribution of threshold credible intervals, $c$, ($\hat{F}(c)$) plot for the final consensus photo-$z$ estimates for the galaxy (or host-dominated AGN) population (top) and the optical, infrared and X-ray selected AGN population (bottom) within each of the deep fields. Coloured lines represent the distributions in bins of apparent optical magnitude ($I$-band for Bootes, $r$-band for EN1 and LH). Lines that fall above the 1:1 relation illustrate under-confidence in the photo-$z$ uncertainties (uncertainties overestimated) while lines under illustrate over-confidence (uncertainties underestimated).}
 \label{fig:uncertainty_validation}
\end{figure*}

In Fig.~\ref{fig:uncertainty_validation} we illustrate the accuracy of the final calibrated redshift posteriors for the AGN and galaxy subsets.
Shown in both plots are the cumulative distribution ($\hat{F}(c)$) of threshold credible intervals, $c$, for bins of apparent optical magnitude (coloured lines).
As outlined in Section~\ref{sec:pz_accuracy_method}, uncertainty calibration was performed as a function of $I$ band for Bo\"{o}tes and the deepest available $r$-band for EN1 and LH.
For all three LoTSS Deep fields, the uncertainties for the galaxy spectroscopic sample are very well calibrated, with the measured uncertainties lying close to the desired 1:1 relation across the full magnitude range.

For the AGN population, the uncertainties are also generally very well calibrated across the available magnitude ranges.
However, in Bo\"{o}tes we find that the wings of our photo-$z$ posteriors overestimate the true uncertainties for the very brightest ($I < 19$) and faintest ($I > 23$) AGN population.
In these regimes the available spectroscopic training sample is much more limited and the optimisation of the uncertainties is therefore still dominated by the more numerous AGN types.
When compared to the results obtained in \citetalias{2019A&A...622A...3D}, we find that the photo-$z$ uncertainties presented in this work are more accurate (with $\hat{F}(c)$ closer to the desired 1:1 trend).
However, we caution that this does not necessarily mean the photo-$z$ posteriors at a given magnitude are more precise.

As with the training of the machine learning estimates, we note that the large spectroscopic training samples available in the forthcoming WEAVE-LOFAR spectroscopic survey will allow for significant improvements in the calibration of the photo-$z$ uncertainties in future studies.
Nevertheless, the analysis presented in this section verifies that the photo-$z$s provided for the LoTSS Deep Field release are of a high quality and suitable for scientific exploitation.
 
\section{Stellar mass estimates}\label{sec:stellarmasses}
Thanks to the extensive panchromatic photometry available in each field (including deblended photometry for mid-IR to FIR), detailed physical properties can be derived for the LOFAR source population through SED fitting. 
Future papers will exploit this information to derive robust source classifications, AGN accretion modes and star-formation rates for the faint radio source population, however running such codes for the full optical catalogues ($>10^{6}$ sources) is impractical.
 Nevertheless, estimating the stellar masses of the full optically selected population is critical for understanding how radio AGN affects galaxy evolution. 
  Here, we present stellar mass estimates (and rest-frame optical colours) using a simpler grid-based SED fitting approach that scales to the massive samples available across the three deep fields. 
      
  \subsection{Spectral energy distribution fitting}
 Stellar masses are estimated using the \textsc{Python}-based SED fitting code previously used by \citet{Duncan:2014gh} and \citet{2019ApJ...876..110D}. 
Composite stellar populations are generated using the stellar population synthesis models of  \citet{Bruzual:2003ckb} for a \citet{2003PASP..115..763C} initial mass function (IMF), with the following assumptions:

\textit{Star-formation histories}: Recent studies have shown that the widely used assumption of exponentially declining star-formation histories can lead to biases in the ages and star-formation histories derived from photometric SED fits, resulting in biased stellar mass estimates \citep[see e.g.][and references therein]{2016ApJ...832...79P,2019ApJ...873...44C,2019ApJ...876....3L}.
\citet{2018MNRAS.480.4379C} show that the more complex double power-law parametrisation provides sufficient flexibility to accurately describe the star-formation histories of a wide range of possible formation and quenching mechanisms.
As a compromise between the tractability of fitting large samples and the optimal prior assumption on star-formation histories, we therefore define a grid of SFH based on the double power-law model with the priors on the range of power-law slopes and turnover ages taken from \citet{2019MNRAS.490..417C}.
Specifically, we use a logarithmically spaced grid of 7 values from $0.1 \leq \alpha_{i} \leq 1000$ for both power-law slopes ($i \in \left \{ 1, 2 \right \}$) and 7 turnover ages from 0.1 to 1 times the age at observation (totalling 343 star-formation history models for every age step). 

For a given source the time since the onset of star-formation is defined as the time since a fixed formation redshift of $z_{\textup{f}} = 20$ at the closest redshift step in a grid from $0 < z < 1.5$. 

\textit{Stellar metallicity}: We assume constant stellar metallicities at fixed values of $Z \in \left \{ 0.1, 0.4, 1.0 \right \} Z_{\odot}$.

\textit{Nebular emission}: Due to the redshift range being probed and the available constraints on the rest-frame NIR SEDs of our target sample, nebular emission lines are not expected to have a significant effect on the inferred stellar masses \citep[unlike at $z > 3$ where they can have a significant effect, e.g.][]{Stark:2013ix, Schenker:2013ep}.
 Nevertheless, a simple prescription for nebular emission is included in the model SEDs allowing for escape fractions of $f_{\textup{esc}} \in \left \{ 0., 0.2 \right \}$. 
 Details of the assumed emission line ratios for Balmer and metal lines, as well as the nebular continuum prescription can be found in \citet{Duncan:2014gh}. As in previous studies we make the simplifying assumption that the gas-phase stellar metallicity equals the stellar metallicity. 

\textit{Dust attenuation}: We incorporate dust attenuation following the two-component dust model of \citet{2000ApJ...539..718C}, which includes contributions from birth clouds (for stellar populations younger than $10^{7}$ years) and an additional screen component for all ages.
The wavelength dependence on the extinction curve and fraction of attenuation from the ISM are fixed at $A(\lambda) \propto \lambda^{-0.7}$ and 0.3, respectively - as originally presented by \citet[][to which we refer the reader for formal definitions]{2000ApJ...539..718C}.\\

After convolution of the model spectral energy distributions with the photometric filters of each field at each redshift in the grid, the model grid is fit to the observational datapoints for all optical catalogue sources with a photo-$z$ $z_{1, \textup{median}}  < 1.5$ -- the regime for which photo-$z$ can be considered reliable for the galaxy population.
Fitting to the observed photometry is done using a simple least-squares fit to all available photometric bands for a given source \citep[see Section 2.5 of][for details]{2019ApJ...876..110D}. 
However, rather than using simply the normalisation (and hence stellar mass) of only the best-fit individual model, we marginalise over the full set of stellar population parameters (star-formation history, dust attenuation, metallicity) to derive a likelihood weighted distribution for the inferred stellar mass.
This marginalisation implicitly assumes a flat prior on the respective stellar population parameters.
For the resulting catalog, we then take the median and 1$\sigma$ (16 and 84th percentiles) of the stellar mass distribution as the estimate and corresponding uncertainty for a given source.
We note however that this uncertainty does not reflect the full statistical uncertainty on the stellar masses as we do not account for the uncertainty in redshift.

In addition to the stellar mass estimate for each source, we also derive the rest-frame magnitude for all photometric filters in a given field (regardless of whether that filter was used during fitting) based on the best-fitting template.
Future papers that incorporate the full panchromatic SED (including deblended 24 to 500$\mu$m observations) will provide robust estimates of the star-formation rates and other key physical properties within the LoTSS Deep Field source population.

\subsection{Flux zero-point and total flux corrections}\label{sec:mass_corrections}
During testing and calibration of the stellar mass estimates, we found that additional flux zeropoint corrections and model uncertainties similar to those used during photo-$z$ estimation are required for accurate and reliable mass estimates \citep[see also][]{2015ApJ...801...20T}.
Offsets to the flux zero-points for stellar mass estimates are calculated following a similar approach to that applied during photo-$z$ template fitting.
The subset of optical sources with robust spectroscopic redshifts are first fit with no flux offsets applied and with no additional errors included to account for model uncertainty.
Corrections to each observed filter are then calculated based on the median ratio between the observed best-fit model fluxes in that band.
The sample is then re-fit incorporating the initial zero-point corrections to derive iterative changes to the corrections, with the process repeated a final time.
After the third iteration of zero-point corrections, we then estimate the residual model uncertainties for the SED fitting model grid following the same method as in Section~\ref{sec:photoz-modelerrors}.
The resulting corrections are included in Appendix~\ref{app:zeropoints} for reference.

When the final derived zero-point corrections are applied during fitting, we observe a significant improvement in the distribution of best-fit $\chi^{2}$ values for the full photometric samples. 
For example in EN1 we find a median reduction per source of $\Delta\chi^{2} = -0.68$, and a reduction in the average $\chi^{2}$ for the whole sample by $\approx 2$.
Furthermore, we also find an increased agreement between the observed stellar mass functions (SMFs) between fields and with those available in the literature.
We note that when fitting stellar masses using the zeropoint corrections derived from the photo-$z$ estimates \citep[specifically the combined template sets of][]{Brown:2014jd,2019MNRAS.489.3351B}, we find no systematic offset in the resulting stellar masses - with a median offset of $\lesssim 0.03\,\text{dex}$, significantly smaller than the typical uncertainty on individual estimates.

 However, due to the use of fixed apertures (and aperture corrections) within the optical photometry catalogues, we find that we underestimate the mass of the most massive galaxy population at lower redshifts (which have sizes significantly larger than those of the faint galaxies or stars used for the aperture corrections).
To estimate corrections from aperture to total fluxes for brighter resolved sources we cross-match the optical photometry in the deep fields to the model-fitting photometry of the Legacy Surveys \citep{2019AJ....157..168D}.
For sources detected in Legacy Surveys imaging, we calculate the ratio between the LOFAR optical catalogue \citepalias{Kondapally2020} aperture corrected flux and the model-fitting photometry based total flux in the available bands ($g$, $r$, $z$ for EN1 and LH and $z$ for Bo\"{o}tes).
After accounting for any small global offset between the two flux measurements based on the median flux ratio for the source types used for aperture corrections, the resulting flux ratio can be used to estimate an approximate aperture to total flux conversion for each source. 
In the faint limit where a source has a S/N less than $5\sigma$ or is undetected in Legacy Surveys imaging, we assume that sources in the LOFAR optical catalogue are unresolved and the aperture corrected fluxes provide a reliable estimate of the total flux.
In the stellar mass catalogue presented in this data release we provide the  stellar mass and rest-frame magnitude values as measured, alongside total flux corrections where available. 


\subsection{Stellar mass completeness}\label{sec:mass_completeness}
\begin{figure}[h!]
\centering
 \includegraphics[width=0.9\columnwidth]{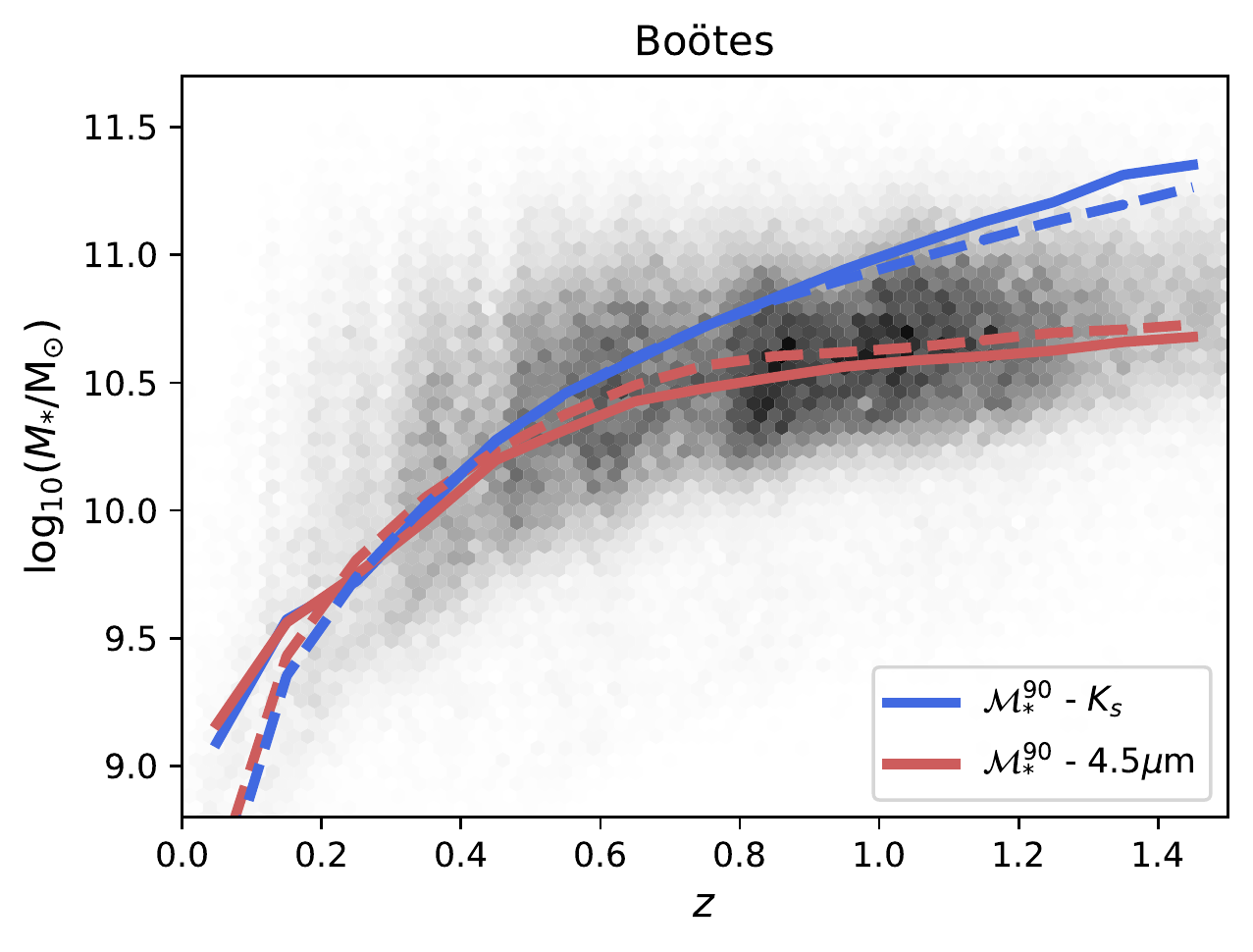}
 \includegraphics[width=0.9\columnwidth]{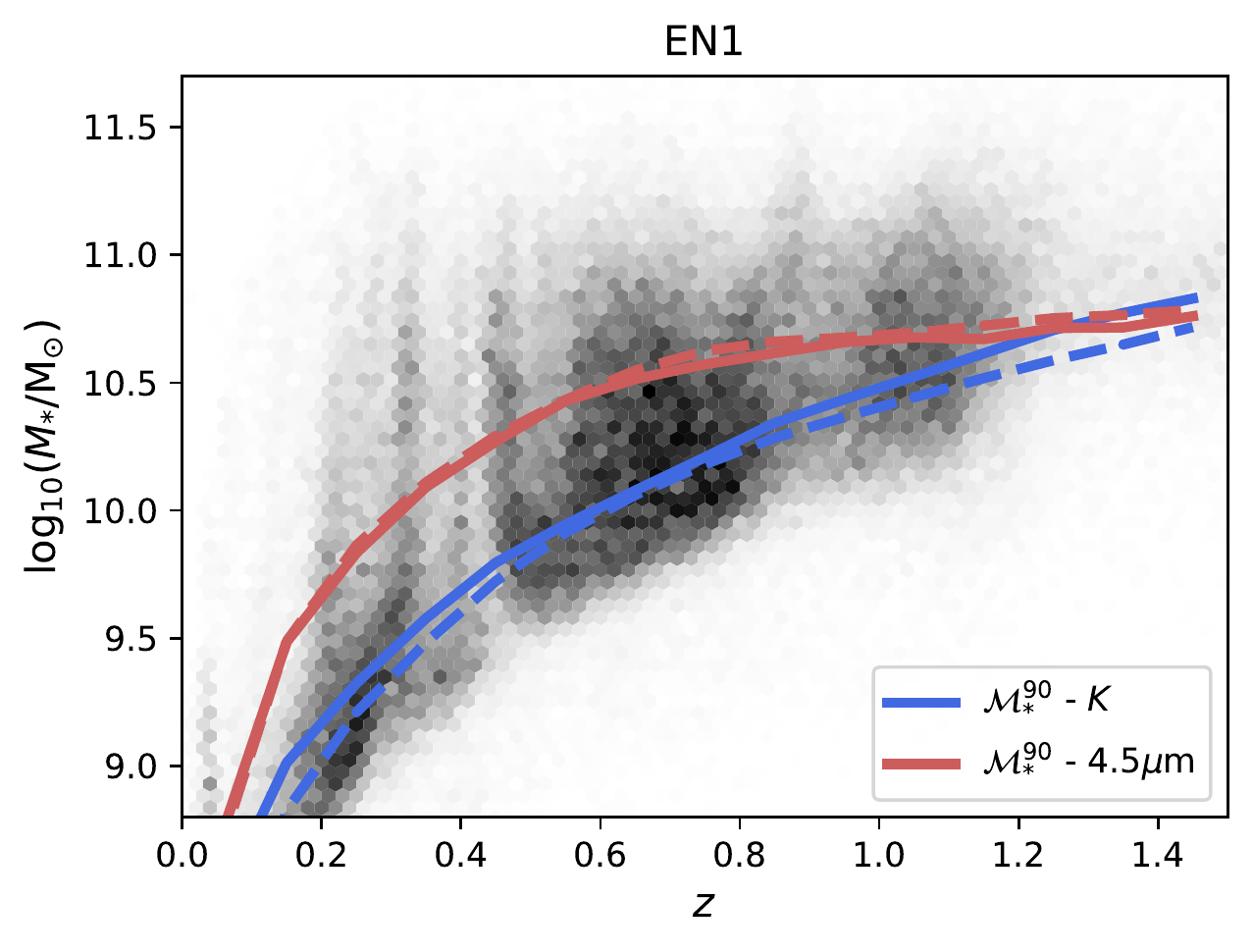}
 \includegraphics[width=0.9\columnwidth]{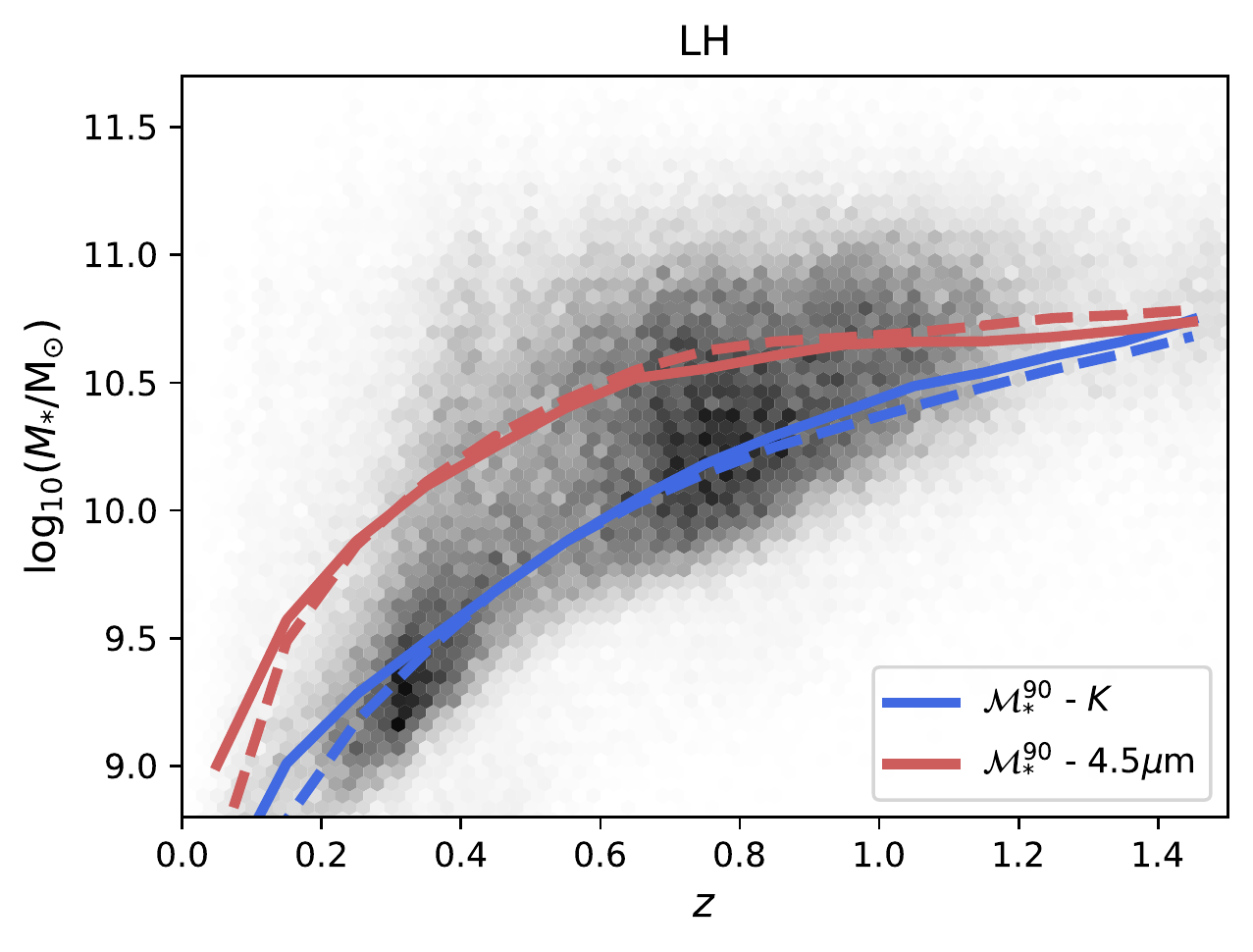}
 \caption{Observed stellar mass distribution as a function of redshift for the three LoTSS Deep Fields. The background density plot shows the mass distribution of sources brighter than the 90\% magnitude limit of the most sensitive reference band available in that field (4.5$\mu$m for Bo\"{o}tes and $K$ for EN1 and LH). Solid lines represent the 90\% mass completeness limits, $\mathcal{M}_{\star}^{90}$, derived empirically \citep{Pozzetti:2010gw} and dashed lines show the corresponding limits derived from model stellar populations. The blue completeness curves are derived based on the $K$ or $K_{s}$ magnitude limit, while red curves are derived from the widest area 4.5$\mu$m limit (i.e. SWIRE for EN1 and LH).}
 \label{fig:mass_completeness}
\end{figure}
When performing studies of different galaxy or AGN populations across redshift and environments in flux limited surveys, understanding the associated completeness limits is essential for minimising biases and defining appropriate volume limited samples.
We are therefore interested in the mass completeness limits of our surveys in order to enable reliable studies in mass-selected samples.

We estimate the detection completeness limit in the $K$ (or $K_{s}$) and IRAC 4.5$\mu$m bands empirically by fitting a power law distribution to the observed number counts for $>5\sigma$ sources in a regime where the catalogues are known to be complete.
The 90\% and 50\% completeness limits are then defined as the magnitude at which the observed number counts are 0.9 and 0.5$\times$ the expected number counts predicted by the power law distributions. 
Although the true magnitude completeness limits will vary as a function of intrinsic source size and morphology, completeness estimates derived from simple power-law fits have been shown to be in good agreement with completeness estimates derived from detailed simulations \citep[through injection and detection of fake sources within the imaging, see e.g.][]{Guo:2013ig}.
We estimate the 90\% magnitude completeness limits to be $K$ or $K{s} = $ 20.44, 21.78, and 21.87 for Bo\"{o}tes, EN1, and LH, respectively.
While for the IRAC 4.5$\mu$m band we estimate 90\% magnitude completeness limits of 21.33, 21.19, and 21.18 respectively, based on the SDWFS observation in Bo\"{o}tes and the SWIRE observations in the other two fields.

To estimate the mass completeness limit associated with the measured detection completeness limit, we derive the limiting stellar mass-to-light ratio as a function of redshift in two ways.
Firstly, following the empirical method presented by \citet{Pozzetti:2010gw}, in small redshift bins we take the 20\% faintest sources above the 90\% magnitude completeness limit.
The measured stellar masses for this sample are scaled to the magnitude limit ($\log_{10}(\mathcal{M}_{\textup{lim}}) = \log_{10}(\mathcal{M}) + 0.4(m - m_{\textup{lim}}^{90})$) and the mass completeness limit derived from 95th percentile of the scaled $\mathcal{M}_{\textup{lim}}$ mass distribution.
This measurement is repeated for both the observed stellar masses and $K$ ($K_{s}$) and IRAC 4.5$\mu$m magnitude limits in each field.

Our second derivation of the mass completeness limit is based on a maximally old stellar population representative of the highest mass-to-light ratio expected within the sample.
Using the same stellar populations models as used for SED fitting, we model a dust-free exponentially declining star-formation history with an $e$-folding time of 50 Myr and a formation redshift of $z=20$.
For each redshift bin, the stellar population is convolved with the $K$ ($K_{s}$) and IRAC 4.5$\mu$m filter response curves with the stellar population age set by the time since the formation redshift.
The mass completeness limit corresponding to the maximally old population is then derived by scaling the observed template flux at each redshift bin to the corresponding magnitude completeness limit.

In Figure~\ref{fig:mass_completeness} we find that the two approaches for estimating the limiting mass-to-light ratio at each redshift yield almost identical results.
Selecting based on either near- or mid-IR, all three fields are complete to $\log_{10}(\mathcal{M}_{\star}^{90}/M_{\odot}) \approx 10.7$ at $1.4 < z \leq 1.5$ - sufficient to probe below the knee of the galaxy SMF at this redshift.
At $z < 1$, the deeper NIR observations provided by UKIDSS Deep Extragalactic Survey (DXS) mean that the EN1 and LH fields are complete to significantly lower masses if using $K$ band to select samples.
We note that by restricting observations to the \emph{Spitzer} SERVS region of the EN1 and LH fields ($\approx 1$ magnitude deeper than SWIRE), mass complete samples at $z \sim 1.5$ could likely reach $\sim0.4\,\textup{dex}$ lower in mass at the expense of total volume probed.

\begin{figure}[h!]
\centering
 \includegraphics[width=0.97\columnwidth]{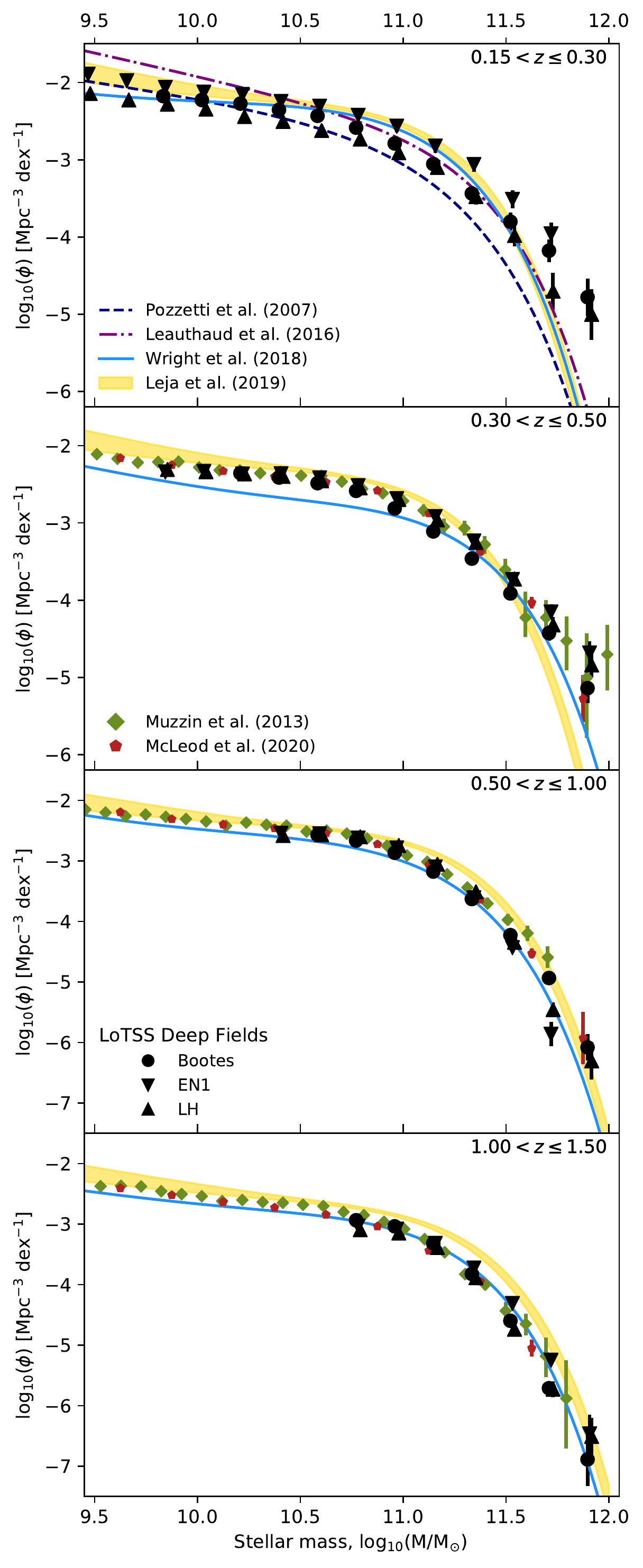}
 \caption{Estimated galaxy SMFs in each LoTSS deep field based on the stellar mass estimates presented in this data release (black data points), with a sample of results from the literature over the same redshift ranges. For the SMF estimates from this work, we only plot only the mass ranges that are estimated to be 90\% complete (see Section~\ref{sec:mass_completeness}. Total flux corrections to the inferred stellar masses are applied for all fields at $z < 0.5$ and for Bo\"{o}tes at $z < 1$.}
 \label{fig:smf_comparison}
\end{figure}

\subsection{Stellar mass validation}
To validate our stellar mass estimates and verify the consistency both between the three LOFAR deep fields and with the literature, we estimate the galaxy SMF in each field.
After masking regions affected by bright stars and restricting to regions with both optical and NIR or mid-IR coverage, the resulting survey areas in each field are 8.63, 6.5 and 7.24 $\textup{deg}^2$ for Bo\"{o}tes, EN1, and LH, respectively (we note that these areas correspond to the region where our selection criteria have been satisfied, rather than the full areas quoted in \citetalias{Kondapally2020}).
For the purposes of this comparison, we construct our estimate of the galaxy SMF through simple calculation of the volume density in bins of stellar mass.
We do not include additional corrections to weight for the reduced volume for sources near the detection limit (i.e. the classical $1/V_{max}$ method) or for the effects of Eddington bias on the high mass bins.

For each redshift range, we limit to the 90\% mass completeness limit based on the deepest available band over a large area, specifically the UKIDSS DXS $K$ for the EN1 and LH fields, and the IRAC $4.5\mu$m band for Bo\"{o}tes.
We exclude known IR, X-ray and optical AGN that are likely to result in poor SED fits or biased stellar mass estimates. 
Additionally, we apply a cut on the best-fit $\chi^{2}$ to exclude sources with $\chi^{2}/N$ (where $N$ equals the number of filters used during fitting) to exclude the 5\% worst fits indicative of unidentified stars or AGN within the sample.
 
Our derived SMF estimates for four redshift bins ($0.15 < z \leq 0.3$, $0.3 < z \leq 0.5$, $0.5 < z \leq 1.0$, and $1.0 < z \leq 1.5$) are presented in Figure~\ref{fig:smf_comparison} alongside a selection of published SMFs from the literature that probe comparable redshifts \citep[][in prep.]{2007A&A...474..443P,Muzzin:2013bl,2016MNRAS.457.4021L,2018MNRAS.480.3491W,2020ApJ...893..111L, McLeod2020}.
We plot error bars for our binned SMF estimates that represent the uncertainty only from Poisson noise and the approximate cosmic variance based on the volume probed for each field.
Our cosmic variance uncertainties are calculated following the prescription and code presented in \citet{Moster:2011ip}, with the standard deviation on the number counts at a given mass interpolated from the mass ranges produced by the code.

 As noted above, we find that total flux corrections are necessary to produce a broad agreement between the three fields and with observed SMFs and those of the literature.
However, due to the different assumptions used for aperture corrections between Bo\"{o}tes (analytic corrections based on the PSF), EN1 and LH (empirical corrections), we include total flux corrections at $z < 0.5$ for the EN1 and LH fields, while for Bo\"{o}tes we include corrections out to $z < 1$. 
We find that corrections are not required at higher redshift, with the three fields providing results that are consistent with the population becoming largely unresolved.

At $z > 0.3$, where the fields probe a large representative volume, our simple galaxy SMFs are in excellent agreement both with published literature values and with self-consistent results across all three fields. 
In the lowest redshift bin, $0.15 < z \leq 0.3$, we observe an increased scatter between the SMF estimates of the three different fields, with EN1 having a consistently higher normalisation than the other two fields.
Given the significantly smaller volume probed in this bin ($\sim 3.4\times$ less than at $0.3 < z \leq 0.5$), larger cosmic variance between fields is to be expected.
While the estimated cosmic variance uncertainties do not fully account for the offset between fields in all redshift bins, the overall normalisation and shape of our observed SMFs are still broadly consistent with the range published in the literature.

Finally, as with the photo-$z$s, we can compare our stellar mass estimates on a source by source basis with those available in the literature.
Specifically, we compare our EN1 masses with those presented in the HSC PDR2 \citep[][see Sec.~\ref{sec:photz_lit_comp}]{2020arXiv200301511N}, which assume the same \citet{2003PASP..115..763C} IMF as used in this work.
For sources with $\log_{10}(\mathcal{M}_{\star}/M_{\odot}) > 9$ in our catalog, we find a median offset of $0.09\,\text{dex}$ between our stellar masses and those of the HSC-DeMP \citep[][]{2014ApJ...792..102H} methodology, with a corresponding robust scatter of $0.28\,\text{dex}$.
For the HSC-\textsc{Mizuki} template based estimates \citep[][]{2015ApJ...801...20T}, we find a median offset of just $0.01\,\text{dex}$ and robust scatter of $0.32\,\text{dex}$.
We note that the typical difference between our estimates and either of the HSC estimates is smaller than the difference between the two HSC estimates themselves when calculated for the same galaxy sample (a median offset of $0.14\,\text{dex}$).

Given the good agreement and consistency between our three fields at the redshifts of interest, we are therefore confident, that with appropriate sample selection and quality cuts, our stellar mass estimates are suitable for robust quantitative studies across all three fields.
Furthermore, the large samples and volumes probed by the combined deep fields dataset over a wide range of cosmic history offer reliable reference samples for studying radio properties as a function of history, mass, and environment.

\clearpage

\section{Radio source properties}\label{sec:radioproperties}
The combination of exquisite radio continuum observations and extensive multi-wavelength data covering $>20$ deg$^2$ provided by the LoTSS Deep Fields data release offers the potential for a wide range of studies of the radio population - ranging from studies of the cosmic star-formation history through to obscuration free studies of the black hole accretion history \citep[e.g.][]{2017A&A...602A...6S,2017A&A...602A...5N}.
In addition to the SED fitting presented in this paper, extensive detailed physical modelling of the radio population that incorporates additional FIR information \citep[e.g.][]{2016ApJ...833...98C} can offer further insight into the physical properties of the faint radio source population.
Building upon the multi-wavelength datasets presented in \citetalias{Kondapally2020} and the photo-$z$s presented here, Best et al. (2020; Paper V) combine results from multiple SED fitting tools to derive consensus source classifications for the radio population. 
Further papers will utilise these classifications and the associated detailed SED modelling to study various open questions regarding AGN and galaxy evolution.

\begin{figure}[]
\centering
 \includegraphics[width=\columnwidth]{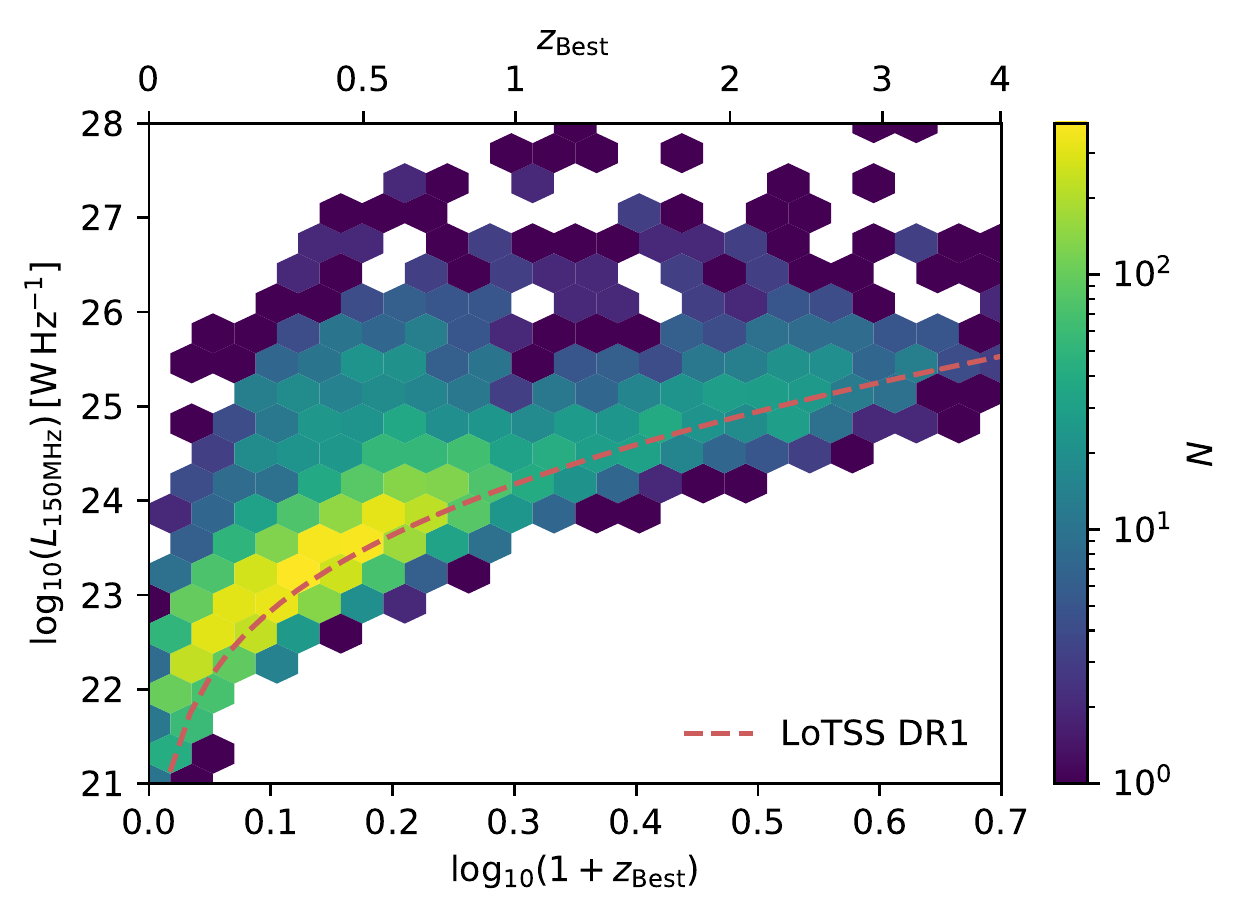}
 \caption{Distribution of LoTSS Deep Field sources as a function of redshift, $z_{\textup{Best}}$ (\texttt{Z\_BEST} in our catalogues), and radio power $L_{150\textup{MHz}}$ assuming a constant spectral index of $\alpha = -0.7$. The colour scale illustrates the combined number of sources across the three deep fields with a logarithmic scale. The 5$\sigma$ luminosity limit for the wide area LoTSS DR1 \citep{Shimwell:2018to} data is plotted in red for comparison.}
 \label{fig:z_vs_power}
\end{figure}

We therefore do not present a detailed analysis of the LoTSS Deep Field radio source population in this study.
However, to demonstrate the scientific potential offered by the LoTSS Deep Fields release, we present some basic properties of the radio source population derived from the catalogues presented in this work.
In Figure~\ref{fig:z_vs_power}, we show the overall distribution of the LoTSS Deep Field sources as a function of both redshift (based on our best available estimate, $z_{\textup{Best}}$, where spectroscopic redshifts are used when available and photo-$z$s otherwise) and of the inferred low-frequency radio luminosity, $L_{150\textup{MHz}}$.

\begin{figure}[]
\centering
 \includegraphics[width=\columnwidth]{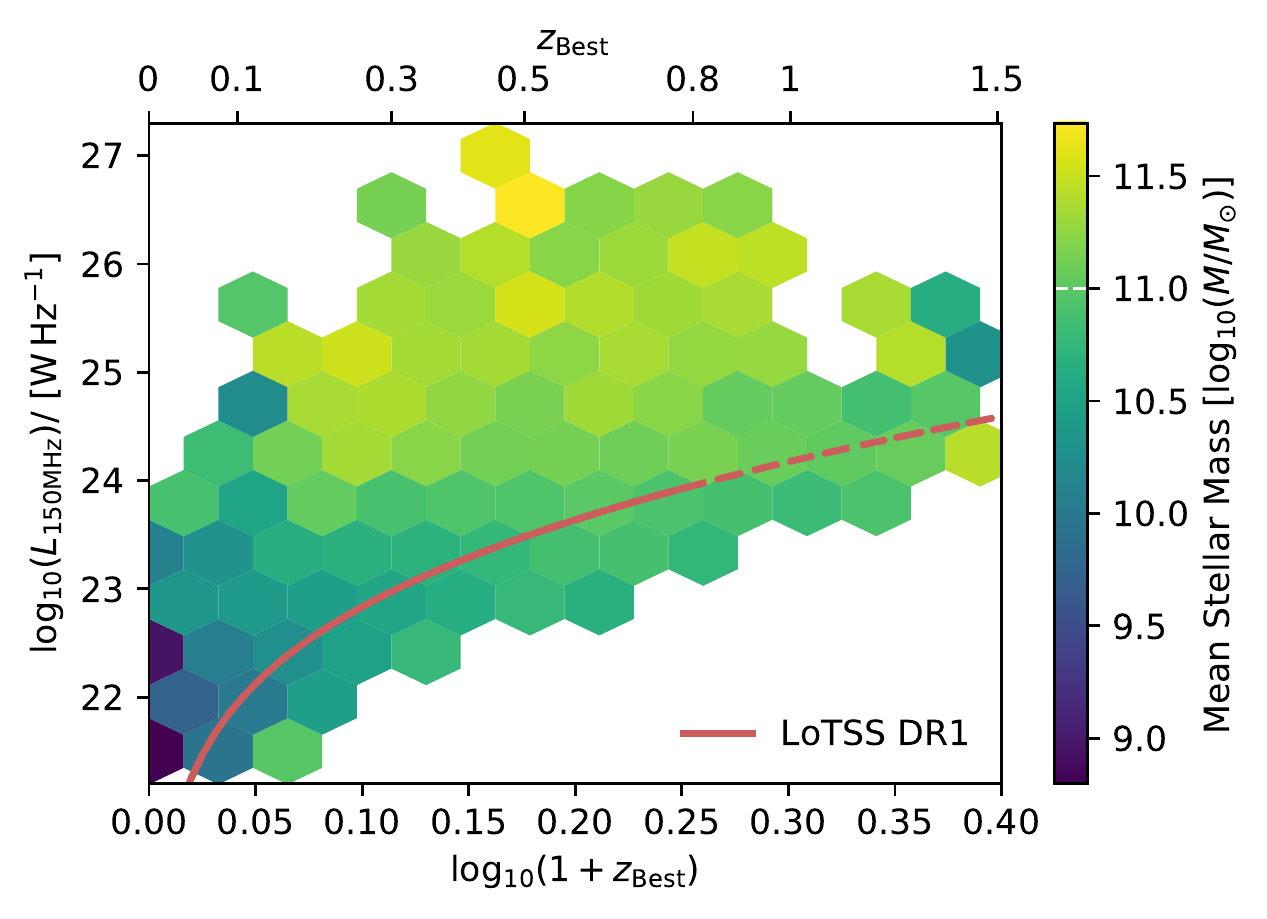}
 \caption{Average stellar mass (taken as the mean in $\log_{10}$ space) of the LoTSS Deep radio population as a function of redshift, $z_{\textup{Best}}$ (\texttt{Z\_BEST} in our catalogues), and radio power $L_{150\textup{MHz}}$ - assuming a constant spectral index of $\alpha = -0.7$. The 5$\sigma$ luminosity limit for the wide area LoTSS DR1 \citep{Shimwell:2018to} data is plotted in red for comparison \citepalias[with the solid line representing the redshift range with high quality photo-$z$ for the comparable galaxy population][]{2019A&A...622A...3D}. The typical $M^{\star}$ for the galaxy SMF at these redshifts is marked on the colour scale for reference (white dashed line).}
 \label{fig:z_vs_power_mass}
\end{figure}

The colour scale in Figure~\ref{fig:z_vs_power} illustrates the total number of sources per cell across all three fields, with no additional correction for radio source completeness.
However, we do apply a cut on the measured photo-$z$, restricting the sample to $z_{\textup{Best}} < 4$ and also exclude sources affected by bright stars.
Clearly illustrated by this figure is the broad dynamic range offered by the LoTSS Deep Fields, which offer large samples of low power radio sources ($\log_{10}(L_{150\textup{MHz}}/ \textup{W~Hz}^{-1}) < 25$) while also probing a sufficient volume to detect statistical samples of the high power radio source population.

In Figure~\ref{fig:z_vs_power_mass}, we present a different view of the radio source population, showing the mean log stellar mass in cells of redshift and radio luminosity.
We limit the analysis to $z_{\textup{Best}} < 1.5$, where stellar mass estimates are available, and exclude sources with identified AGN components likely to bias the stellar mass estimates.
At low redshift and low radio power, the average stellar mass is significantly below $M^{\star}$ at these redshift ranges ($M^{\star}\approx 10^{11} \text{M}_{\odot}$, see Figure~\ref{fig:smf_comparison}); this is consistent with the expectation that the radio continuum population at these luminosities is dominated by star-forming galaxies \citep{2005MNRAS.362...25B, 2017A&A...602A...2S}.
As radio luminosity increases, the radio population is hosted by increasingly massive galaxies.
This apparent evolution is likely driven by two trends: the first being the transition from the regime where star-forming galaxies dominate the radio source population to where AGN dominate, and the second being the strong correlation between stellar mass and radio AGN activity for radio AGN \citep[][]{2019A&A...622A..17S}.

We note the important caveat that this analysis excludes the radio AGN that are hosted by sources with significant dust torus emission (as would be detected by IRAC colour criteria) as well as radio-quiet quasars. 
The trends observed are therefore not representative of the full evolution in radio AGN host properties over this redshift range.
Nevertheless, Figure~\ref{fig:z_vs_power_mass} illustrates the potential of the catalogues presented in this work, which when combined with additional source classification and complex SED fitting can offer a detailed picture of the evolution of the radio source population.
                                                                 
\section{Summary}\label{sec:summary}

In this paper we present details of photo-$z$ and stellar mass estimates produced for the LoTSS Deep Fields DR1.
Photo-$z$s are estimated for all optical sources within the three Deep Fields (Bo\"{o}tes, ELAIS-N1 and LH), totalling over 5 million estimates across a combined $\sim25
~\textup{deg}^2$ after appropriate optical quality cuts.
Building on previous work our photo-$z$ method combines multiple template fitting and empirical training based estimates to produce a consensus redshift prediction with well-calibrated photo-$z$ uncertainties.

Based on the available spectroscopic training and test sample in each field, the resulting consensus photo-$z$s have robust scatter ranging from $\sigma_{\textup{NMAD}} =  0.016$ to 0.02 for galaxies and/or host dominated AGN sources, and from $\sigma_{\textup{NMAD}} =  0.064$ to 0.07 for identified AGN sources.
Our estimated OLF for the corresponding subsets range from 1.5 to 1.8\% and 18 to 22\%, respectively.
Similar to previous studies we find that the photo-$z$ quality is a function of both optical magnitude and spectroscopic redshift.

Exploring the photo-$z$ quality of the LoTSS Deep Field radio source populations we find that there is no strong trend in photo-$z$ quality as a function of radio luminosity (for a fixed redshift), reproducing trends observed in previous studies \citep{2019A&A...622A...3D}.
However, there is clear deterioration in photo-$z$ quality as a function of redshift for a given radio luminosity that we attribute to selection effects in the spectroscopic sample and/or intrinsic evolution within the radio population.

In the redshift range for which we find our photo-$z$ estimates to be reliable for host-dominated SEDs ($z < 1.5$), we exploit the extensive wavelength coverage to produce consistent stellar mass estimates across the three fields, as well as estimates of the corresponding stellar mass completeness.
From simple estimations of the galaxy SMFs within each field and comparison with the literature, we validate our stellar masses provide reliable and self consistent estimates suitable for statistical studies across all three fields.

The catalogue presented in this work builds upon both the LOFAR radio data presented in Paper I \& II, and the optical catalogues and radio-optical cross identification presented in \citetalias{Kondapally2020}.
All data produced in this work are made available for public release to enable the full exploitation of the LoTSS Deep Field survey for a wide range of scientific goals.

\section*{Acknowledgements}
KJD and HJAR acknowledge support from the ERC Advanced Investigator programme NewClusters 321271.
RK acknowledges support from the Science and Technology Facilities Council (STFC) through an STFC studentship via grant ST/R504737/1.
PNB and JS are grateful for support from the UK STFC via grant ST/R000972/1
MB acknowledges support from INAF under PRIN SKA/CTA FORECaST and from the Ministero degli Affari Esteri della Cooperazione Internazionale - Direzione Generale per la Promozione del Sistema Paese Progetto di Grande Rilevanza ZA18GR02.
RAAB acknowledges support from the Glasstone Foundation.
MJH acknowledges support from the UK STFC via grant ST/R000905/1.
MJJ acknowledges support from the UK Science and Technology Facilities Council [ST/N000919/1] and the Oxford Hintze Centre for Astrophysical Surveys which is funded through generous support from the Hintze Family Charitable Foundation.
MKB and AW acknowledge support from the National Science Centre, Poland under grant no. 2017/26/E/ST9/00216.
K.M. has been supported by the National Science Centre (UMO-2018/30/E/ST9/00082).
IP acknowledges support from INAF under the SKA/CTA PRIN "FORECaST" and the PRIN MAIN STREAM "SAuROS" projects.

This paper is based (in part) on data obtained with the International LOFAR Telescope (ILT) under project codes LC0 015, LC2 024, LC2 038, LC3 008, LC4 008, LC4 034 and LT10 01. 
LOFAR \citep{vanHaarlem:2013gi} is the Low Frequency Array designed and constructed by ASTRON. It has observing, data processing, and data storage facilities in several countries, which are owned by various parties (each with their own funding sources), and which are collectively operated by the ILT foundation under a joint scientific policy.
The ILT resources have benefitted from the following recent major funding sources: CNRS-INSU, Observatoire de Paris and Universit\'{e} d'Orl\'{e}ans, France; BMBF, MIWF-NRW, MPG, Germany; Science Foundation Ireland (SFI), Department of Business, Enterprise and Innovation (DBEI), Ireland; NWO, The Netherlands, The Science and Technology Facilities Council, UK; Ministry of Science and Higher Education, Poland.

The Pan-STARRS1 Surveys (PS1) and the PS1 public science archive have been made possible through contributions by the Institute for Astronomy, the University of Hawaii, the Pan-STARRS Project Office, the Max-Planck Society and its participating institutes, the Max Planck Institute for Astronomy, Heidelberg and the Max Planck Institute for Extraterrestrial Physics, Garching, The Johns Hopkins University, Durham University, the University of Edinburgh, the Queen's University Belfast, the Harvard-Smithsonian Center for Astrophysics, the Las Cumbres Observatory Global Telescope Network Incorporated, the National Central University of Taiwan, the Space Telescope Science Institute, the National Aeronautics and Space Administration under Grant No. NNX08AR22G issued through the Planetary Science Division of the NASA Science Mission Directorate, the National Science Foundation Grant No. AST-1238877, the University of Maryland, Eotvos Lorand University (ELTE), the Los Alamos National Laboratory, and the Gordon and Betty Moore Foundation.

This work is based on observations obtained with MegaPrime/MegaCam, a joint project of CFHT and CEA/DAPNIA, at the Canada-France-Hawaii Telescope (CFHT) which is operated by the National Research Council (NRC) of Canada, the Institut National des Sciences de l'Univers of the Centre National de la Recherche Scientifique (CNRS) of France and the University of Hawaii.
This research used the facilities of the Canadian Astronomy Data Centre operated by the National Research Council of Canada with the support of the Canadian Space Agency.
RCSLenS data processing was made possible thanks to significant computing support from the NSERC Research Tools and Instruments grant program. 
This work is based in part on observations made with the Spitzer Space Telescope, which was operated by the Jet Propulsion Laboratory, California Institute of Technology under a contract with NASA.

\bibliographystyle{aa}
\bibliography{bibtex_library}
\clearpage
\newpage

\begin{appendix}
\section{Catalogue description}\label{sec:catalog}
The contents of the catalogue added by this work are as follows:
\begin{itemize}
\item  \texttt{Z\_BEST} - Best available redshift estimate
\item  \texttt{Z\_BEST\_SOURCE} - The source of the best available redshift, \texttt{Z\_BEST} where 1 corresponds to spectroscopic redshift and 0 corresponds to the photo-$z$ presented in this work.

\item  \texttt{Z\_SPEC} - Literature Spectroscopic Redshift
\item  \texttt{Z\_SOURCE} - Source of the spectroscopic redshift
\item  \texttt{Z\_QUAL} - Spectroscopic redshift quality, where flag $Q = 3$ means probable, $Q \geq 4$ means reliable. Lower reliability redshifts have not been included.
\item \texttt{AGN\_ZSPEC} - Spectroscopic AGN/QSO flag where provided
\item \texttt{z1\_median} - Median of the primary redshift peak above 80\% HPD CI
\item \texttt{z1\_min} - Lower bound of the primary 80\% HPD CI peak|
\item \texttt{z1\_max} - Upper bound of the primary 80\% HPD CI peak
\item \texttt{z1\_area} - Integrated area of the primary 80\% HPD CI peak
\item \texttt{z2\_median} - Median of the secondary redshift peak (if present) above 80\% HPD CI
\item \texttt{z2\_min} - Lower bound of the secondary 80\% HPD CI peak
\item \texttt{z2\_max} - Upper bound of the secondary 80\% HPD CI peak
\item \texttt{z2\_area} - Integrated area of the secondary 80\% HPD CI peak
\item \texttt{nfilt\_eazy} - Number of filters included in EAZY template fit|
\item \texttt{nfilt\_atlas} - Number of filters included in Atlas+AGN template fit
\item \texttt{nfilt\_ananna} - Number of filters included in Ananna et al. template fit
\item \texttt{chi\_r\_best} - $\chi^2$ / nfilt for best-fit galaxy or AGN template (any library)
\item \texttt{chi\_r\_stellar} - $\chi^2$ / nfilt for best-fit stellar template
\item \texttt{stellar\_type} - Stellar type of best-fit stellar template
\end{itemize}

Also included for all sources are the multi-wavelength AGN classifications used during photo-$z$ estimation.
\begin{itemize}
\item \texttt{AGN} - Sources flagged by any one of optAGN, IRAGN or XrayAGN
\item \texttt{optAGN} - Flag indicating whether source is included in Million Quasar catalogue compilation \citep{2015PASA...32...10F}, where 1 means a source is included. Sources flagged as AGN based on their spectroscopic redshifts are also flagged.
\item \texttt{IRAGN} - Source satisfies \citet{Donley:2012ji} IR AGN selection criteria.
\item \texttt{XrayAGN} - Source has X-ray counterpart.
\end{itemize}

For the Bo\"{o}tes field, where sources have been matched to the X-B\"{o}otes \emph{Chandra} survey of NDWFS \citep{Kenter:2005gj}, we provide the additional associated values:
\begin{itemize}
\item \texttt{XrayFlux\_0.5-2}  [$10^{-14}$ erg/cm$^{2}$/s] XBo\"{o}tes Soft X-ray Flux
\item \texttt{XrayHardness} - XBo\"{o}tes X-ray Hardness Ratio
\end{itemize}

%

For sources with either \texttt{Z\_SPEC} or \texttt{z1\_median} $< 1.5$, we provide our estimate of the galaxy stellar mass as well as the rest-frame magnitudes for the individual best-fitting SED model.
The catalogue columns associated with the additonal SED fitting are:
\begin{itemize}
\item \texttt{zmodel} - Model grid redshift used in stellar mass fit
\item \texttt{chi\_best} - $\chi^2$ for the single best-fit model SED (for  \texttt{Z\_BEST})
\item \texttt{Mass\_median} - 50th percentile of the marginalised stellar mass posterior (for  \texttt{Z\_BEST}), in units of $\log_{10}(M/M_\odot)$ 
\item \texttt{Mass\_l68} - 16th percentile of the marginalised stellar mass posterior (for  \texttt{Z\_BEST}), in units of $\log_{10}(M/M_\odot)$ 
\item \texttt{Mass\_u68} - 84th percentile of the marginalised stellar mass posterior (for  \texttt{Z\_BEST}), in units of $\log_{10}(M/M_\odot)$ 
\item \texttt{Nfilts} - Number of photometric bands included in stellar mass fit.
\item \texttt{ap\_to\_model\_[\emph{x}]}  - Estimated to aperture to total flux correction derived from the model fitting photometry of the Legacy Survey in a band, $x$.\footnote{For EN1 and LH, we provide corrections for the $g$, $r$, and $z$ bands calculated using either PS1 or RCSLens photometry, respectively. In Bo\"{o}tes we provide corrections based on the $z_{\textup{Subaru}}$ and $z$ bands.}
\item \texttt{ap\_to\_model\_err\_[\emph{x}]} - Statistical uncertainty on  \texttt{ap\_to\_model\_[\emph{x}]} derived from the combined flux uncertainties of the LoTSS Deep Field optical and Legacy Surveys flux measurement.
\end{itemize}

Finally, for each photometric band used during stellar mass estimation we provide the rest-frame magnitude in that filter for the best-fit SED template: 
\begin{itemize}
\item \texttt{[\emph{x}]\_rest} Rest-frame magnitude in a given band, $x$, for best-fit SED (for  \texttt{Z\_BEST})  
\end{itemize}                

\clearpage

%
%

\section{Zeropoint flux corrections}\label{app:zeropoints}
The inclusion of the zero-point offsets during template fitting has been demonstrated to lead to substantial improvement in photo-$z$ quality.
As outlined in Section~\ref{sec:method-templates}, zero-point offsets for the photo-$z$ estimates for each template library and dataset were derived following the method outlined in \citetalias{Duncan:2017wu}.
Similarly, zero-point corrections for stellar mass estimates are calculated following the method described in Section~\ref{sec:mass_corrections}.
Here we provide the multiplicative correction factors for all template sets in each of the LoTSS Deep Fields.

\begin{table}[h]
\caption{Flux zero-point corrections applied to the Bo\"{o}tes photometry during photo-$z$ fitting or stellar mass estimates. Values quoted are multiplicative corrections applied to the observed fluxes and flux uncertainties. The three photo-$z$ template sets consist of the `EAZY' \citep{Brammer:2008gn}, `Brown' \citep{Brown:2014jd,2019MNRAS.489.3351B} and `Ananna' \citep{2017ApJ...850...66A} libraries as described in Section~\ref{sec:method-templates}. If a filter was not included in the SED fitting for a specific template set, we do not provide a corresponding value.}\label{}
\begin{tabular}{lcccc}
\hline
\multicolumn{5}{c}{Bo\"{o}tes}\\
\hline
Filter & \multicolumn{4}{c}{Template Set} \\
\hline
 & EAZY & Brown & Ananna & Masses \\
\hline
$u$ & 0.979 & 0.957 & 0.946 & 0.999 \\
$B_{W}$ & 1.027 & 1.029 & 1.0 & 0.95 \\
$R$ & 0.986 & 0.972 & 0.97 & 0.882 \\
$I$ & 0.969 & 0.976 & 0.958 & 0.898 \\
$z$ & 0.933 & 0.945 & 0.927 & 0.879 \\
$z_{\text{Subaru}}$ & 1.028 & 1.038 & 1.02 & 0.917 \\
$Y$ & 1.003 & 1.008 & 0.992 & 0.971 \\
$J$ & 0.99 & 0.978 & 0.99 & 0.991 \\
$H$ & 1.044 & 1.053 & 1.096 & 1.076 \\
$K$ & 0.797 & 0.815 & 0.856 & 0.895 \\
$K_{s}$ & 1.004 & 1.024 & 1.061 & 1.082 \\
$3.6\mu$m & 1.002 & 0.989 & 1.0 & 1.079 \\
$4.5\mu$m & 1.058 & 0.994 & 1.006 & 1.015 \\
$5.8\mu$m & 0.99 & 0.958 &  & 1.0 \\
$8.0\mu$m & 0.98 & 0.922 &  & 1.0 \\
\hline
\end{tabular}
\end{table}

\begin{table}
\caption{Flux zero-point corrections applied to the EN1 photometry during photo-$z$ fitting or stellar mass estimates. Values quoted are multiplicative corrections applied to the observed fluxes and flux uncertainties. The three photo-$z$ template sets consist of the `EAZY' \citep{Brammer:2008gn}, `Brown' \citep{Brown:2014jd,2019MNRAS.489.3351B} and `Ananna' \citep{2017ApJ...850...66A} libraries as described in Section~\ref{sec:method-templates}. If a filter was not included in the SED fitting for a specific template set, we do not provide a corresponding value.}
\begin{tabular}{lcccc}
\hline
\multicolumn{5}{c}{EN1}\\
\hline
Filter & \multicolumn{4}{c}{Template Set} \\
\hline
 & EAZY & Brown & Ananna & Masses \\
\hline
\hline
 $u$ & 1.023 & 1.033 & 1.0 & 1.047 \\
$g$ & 1.052 & 1.041 & 1.02 & 0.918 \\
$r$ & 1.048 & 1.024 & 1.0 & 0.88 \\
$i$ & 1.019 & 1.014 & 0.993 & 0.879 \\
$z$ & 0.998 & 1.006 & 0.993 & 1.092 \\
$y$ & 0.992 & 1.007 & 0.994 & 0.738 \\
$g_{\text{HSC}}$ & 0.951 & 0.945 & 0.926 &  \\
$r_{\text{HSC}}$ & 0.972 & 0.95 & 0.929 &  \\
$i_{\text{HSC}}$ & 0.992 & 0.987 & 0.97 &  \\
$z_{\text{HSC}}$ & 0.933 & 0.944 & 0.93 &  \\
$y_{\text{HSC}}$ & 0.94 & 0.955 & 0.94 &  \\
$\text{NB921}_{\text{HSC}}$ & 0.925 & 0.939 & 0.925 &  \\
$J$ & 1.203 & 1.182 & 1.165 & 1.129 \\
$K$ & 1.088 & 1.061 & 1.087 & 1.072 \\
$3.6\mu$m SERVS & 0.944 & 0.946 & 0.946 & 0.952 \\
$4.5\mu$m SERVS & 1.0 & 0.96 & 0.98 & 0.917 \\
$3.6\mu$m SWIRE & 0.955 & 0.953 & 0.956 & 0.959 \\
$4.5\mu$m SWIRE & 1.014 & 0.961 & 0.978 & 0.951 \\
$5.8\mu$m SWIRE & 1.0 & 1.0 &  & 1.0 \\
$8.0\mu$m SWIRE & 1.16 & 1.057 &  & 1.0 \\
\hline
\end{tabular}
\end{table}

\begin{table}
\caption{Flux zero-point corrections applied to the LH photometry during photo-$z$ fitting or stellar mass estimates. Values quoted are multiplicative corrections applied to the observed fluxes and flux uncertainties. The three photo-$z$ template sets consist of the `EAZY' \citep{Brammer:2008gn}, `Brown' \citep{Brown:2014jd,2019MNRAS.489.3351B},and `Ananna' \citep{2017ApJ...850...66A} libraries as described in Section~\ref{sec:method-templates}. If a filter was not included in the SED fitting for a specific template set, we do not provide a corresponding value.}
\begin{tabular}{lcccc}
\hline
\multicolumn{5}{c}{LH}\\
\hline
Filter & \multicolumn{4}{c}{Template Set} \\
\hline
 & EAZY & Brown & Ananna & Masses \\
\hline
\hline
$u$ & 1.115 & 1.102 & 1.107 & 1.121 \\
$g$ & 0.989 & 0.978 & 0.94 & 0.904 \\
$r$ & 1.008 & 0.987 & 0.946 & 0.911 \\
$z$ & 0.99 & 1.006 & 0.992 & 0.981 \\
$g_{\text{RCS}}$ & 0.921 & 0.915 & 0.879 & 0.873 \\
$r_{\text{RCS}}$ & 0.963 & 0.947 & 0.904 & 0.889 \\
$i_{\text{RCS}}$ & 0.931 & 0.941 & 0.906 & 0.9 \\
$z_{\text{RCS}}$ & 0.919 & 0.935 & 0.921 & 0.927 \\
$J$ & 1.117 & 1.11 & 1.08 & 1.16 \\
$K$ & 1.113 & 1.071 & 1.109 & 1.126 \\
$3.6\mu$m SERVS & 0.962 & 0.952 & 0.969 & 0.933 \\
$4.5\mu$m SERVS & 1.0 & 0.94 & 1.0 & 0.908 \\
$3.6\mu$m SWIRE & 1.0 & 0.977 & 1.0 & 0.954 \\
$4.5\mu$m SWIRE & 1.019 & 0.967 & 1.0 & 0.941 \\
$5.8\mu$m SWIRE & 1.04 & 0.977 &  & 1.0 \\
$8.0\mu$m SWIRE & 1.2 & 1.039 &  & 1.0 \\
\hline
\end{tabular}
\end{table}

\end{appendix}

\label{lastpage}
\end{document}